\documentclass[aps,prx,twocolumn,showpacs,superscriptaddress,amsmath,amssymb]{revtex4-1}
\usepackage{graphicx}
\usepackage{latexsym}
\usepackage{amsmath,amssymb}
\usepackage{bm} 
\usepackage{color}
\usepackage{epsfig}

\usepackage{graphicx}
\usepackage[colorlinks=true,linkcolor=blue,citecolor=blue,urlcolor=blue]{hyperref}
\usepackage{bbold}
\usepackage{gensymb}

\usepackage{pgfplots,pgfplotstable}
\tikzset{every axis plot post/.append style={solid, thin},every mark/.append style={solid,scale=1}}
\usetikzlibrary{decorations.markings}
\tikzset{mdar/.style ={decoration={markings,mark={at position 0.5 with {\fill (2pt,0)--(-2pt,2pt)--(-2pt,-2pt)--cycle;}}},postaction={decorate}}}
\tikzstyle{dsh}=[dash pattern=on 2.5pt off 1.5pt]
\tikzset{intcur/.style ={decorate,decoration={snake,amplitude=.5mm,segment length=3mm}}}
\tikzstyle{zba}=[dash pattern=on 1pt off 1pt,line width=2pt]

\usepackage{simplewick}

\definecolor{myblue}{rgb}{.93, .93, 1}

\def \a {\alpha}
\def \b {\beta}
\def \d {\delta}
\def \D {\Delta}
\def \e {\epsilon}
\def \ve {\varepsilon}

\def \g {\gamma}
\def \G{\Gamma}
\def \k {\kappa}
\def \l {\lambda}
\def \L {\Lambda}
\def \o {\omega}

\def \t {\theta}

\def \s {\sigma}

\def \dag {\dagger}

\def \p {\partial}

\def \apx {\approx}

\def \til {\tilde}

\def \dag {\dagger}

\newcommand{\intv}[1]{\int_{\mbf #1}}

\def \prll {\parallel}
\def \uar {\uparrow}
\def \dar {\downarrow}

\def \rar {\rightarrow}

\def \la {\langle}
\def \ra {\rangle}
\def \fr {\frac}
\def \lf {\left}
\def \ri {\right}

\newcommand{\epvl}[1]{\la#1\ra}

\def \sst {\substack}
\def \Tr {\mathrm{Tr}}
\def \bece {\begin{center}}
\def \ence {\end{center}}
\def \beeq {\begin{equation}}
\def \eneq {\end{equation}}
\def \beal {\begin{aligned}}
\def \enal {\end{aligned}}
\def \bega {\begin{gathered}}
\def \enga {\end{gathered}}
\def \benu {\begin{enumerate}}
\def \ennu {\end{enumerate}}
\def \beit {\begin{itemize}}
\def \enit {\end{itemize}}
\def \bede {\begin{description}}
\def \ende {\end{description}}
\def \betb {\begin{tabular}}
\def \entb {\end{tabular}}
\def \bear {\begin{array}}
\def \enar {\end{array}}

\def \mbf {\mathbf}
\def \mbb {\mathbb}

\def \mca {\mathcal}

\def \bsb{\boldsymbol}
\def \txt {\text}

\newcommand{\comment}[1]{}

\newcommand{\bsub}{\begin{subequations}}
\newcommand{\esub}{\end{subequations}}

\begin{document}


\title{Parquet renormalization group analysis of weak-coupling instabilities with multiple high-order Van Hove points inside the Brillouin zone}

\author{Yu-Ping Lin}
\affiliation{Department of Physics, University of Colorado, Boulder, Colorado 80309, USA}
\author{Rahul M. Nandkishore}
\affiliation{Department of Physics, University of Colorado, Boulder, Colorado 80309, USA}
\affiliation{Center for Theory of Quantum Matter, University of Colorado, Boulder, Colorado 80309, USA}

\date{\today}

\begin{abstract}
We analyze the weak-coupling instabilities that may arise when multiple high-order Van Hove points are present inside the Brillouin zone. The model we consider is inspired by twisted bilayer graphene, although the analysis should be more generally applicable. We employ a parquet renormalization group analysis to identify the leading weak-coupling instabilities, supplemented with a Ginzburg-Landau treatment to resolve any degeneracies. Hence we identify the leading instabilities that can occur from weak repulsion with the power-law divergent density of states. Five correlated phases are uncovered along distinct stable fixed trajectories, including $s$-wave ferromagnetism, $p$-wave chiral/helical superconductivity, $d$-wave chiral superconductivity, $f$-wave valley-polarized order, and $p$-wave polar valley-polarized order. The phase diagram is stable against band deformations which preserve the high-order Van Hove singularity. 
\end{abstract}

\maketitle

\section{Introduction}

Two-dimensional (2D) multilayer moir\'e heterostructures constitute a major platform of modern condensed matter research. These systems manifest enlarged moir\'e superlattices and according nearly flat bands, thereby enjoy remarkably high experimental tunability with interlayer twist angle, vertical gating electric field, external magnetic field, and pressure. A main family of research on moir\'e heterostructures focuses on the twisted bilayer graphene, where the moir\'e flat bands can develop at small twist angles \cite{bistritzer11pnas,kim16nl,yuan18prb,koshino18prx,kang18prx,po18prx}. Unconventional superconductivity, insulating states, and other correlated phases have been observed experimentally, either at a `magic angle' or under certain setup of the other conditions \cite{cao18n1,cao18n2,yankowitz19sci,kerelsky19n,cao20prl,sharpe19s,choi19np,xie19n,lu19n,jiang19n,serlin20s,stepanov20n,saito20np,cao20ax}. The properties of underlying band structures and the origins of correlated phases have attracted enormous interest.

\begin{figure}[b]
\centering
\includegraphics[scale = 1.02]{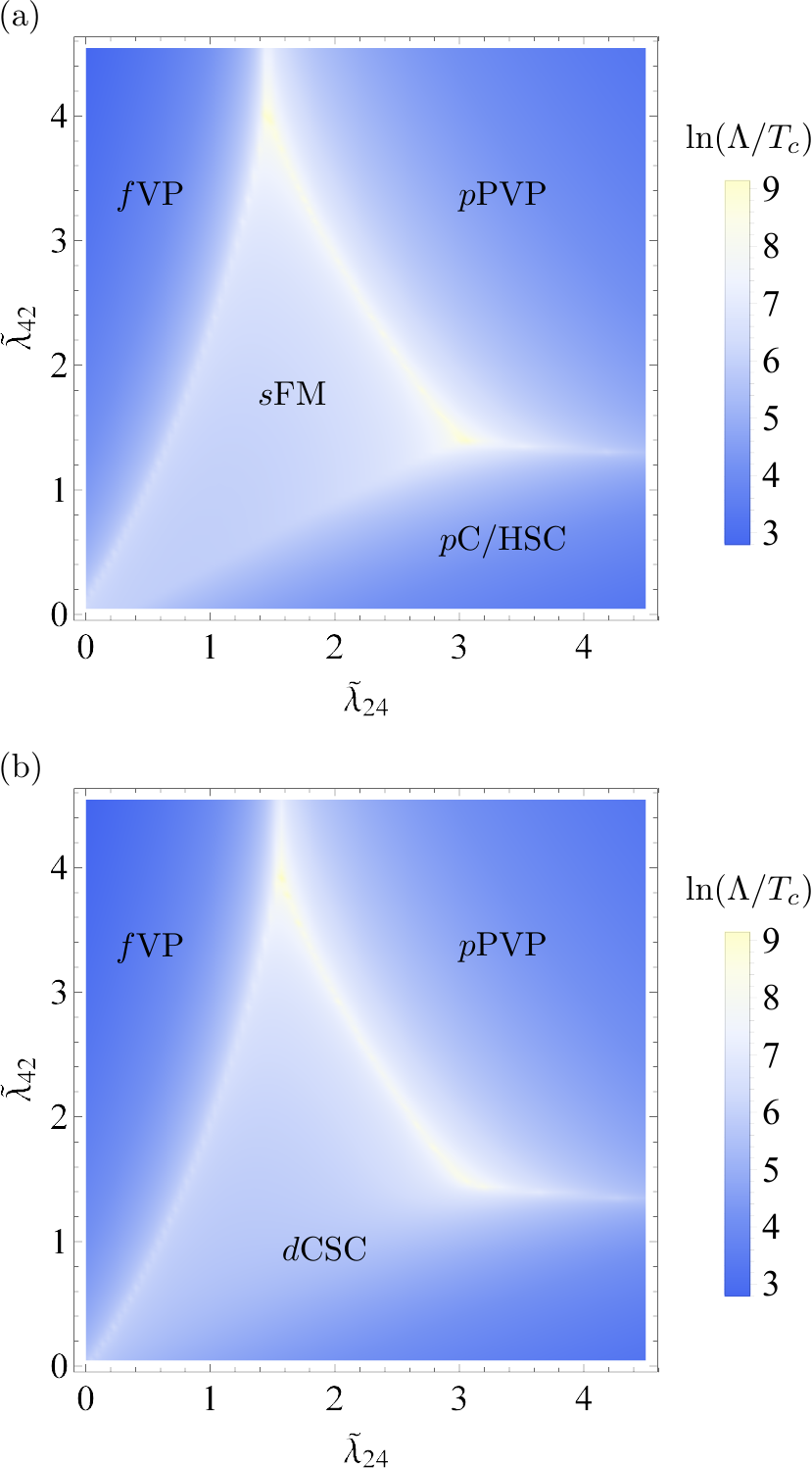}
\caption{\label{fig:pdwive} Tentative electronic phase diagram at the high-order Van Hove singularity in twisted bilayer graphene. The primary interactions are weakly repulsive, with the setup the same as in Fig.~\ref{fig:pd}. Perturbatively (a) repulsive and (b) attractive intervalley exchange interactions are introduced as in Fig.~\ref{fig:rgwive}. Five potential correlated phases are uncovered, including $s$-wave ferromagnetism ($s$FM), $p$-wave chiral/helical superconductivity ($p$C/HSC), $d$-wave chiral superconductivity ($d$CSC), $f$-wave valley-polarized order ($f$VP), and $p$-wave polar valley-polarized order ($p$PVP).}
\end{figure}

For the correlated phases in twisted bilayer graphene, one popular scenario emphasizes the importance of the Van Hove singularity in the density of states \cite{isobe18prx,sherkunov18prb,lin19prb,hsu20prb,liu118prl,kennes18prb,you19npjqm,lin18prb,kozii19prb,classen19prb,chichinadze20prb1,lu20ax,chichinadze20prb2,yuan19nc}. Van Hove singularity can occur at the saddle points of the dispersion energy with divergent density of states \cite{vanhove53pr}. These saddle points are generically present in the dispersive moir\'e flat bands of twisted bilayer graphene \cite{cao18n1,kerelsky19n,choi19np,xie19n,jiang19n}. The model calculations show that the Van Hove singularity occurs near the half-fillings in both electron and hole branches \cite{kim16nl,yuan18prb,koshino18prx,kang18prx,po18prx}. Meanwhile, the corrections from Coulomb interaction may pin the Van Hove singularity to the Fermi surface at a broad range of doping \cite{cea19prb,rademaker19prb}. With the divergent density of states, the electronic correlations can receive significant amplification. Instabilities to the Fermi liquid may occur accordingly, with the energy scales remarkably enhanced compared to the conventional exponentially small ones. By introducing the weakly repulsive interactions at the Van Hove singularity, the investigations of potential correlated phases in twisted bilayer graphene have constituted an enormous literature \cite{isobe18prx,sherkunov18prb,lin19prb,hsu20prb,liu118prl,kennes18prb,you19npjqm,lin18prb,kozii19prb,classen19prb,chichinadze20prb1,lu20ax,chichinadze20prb2}. These works address the interacting problems at the conventional Van Hove singularity, where logarithmically divergent density of states and Fermi surface nesting are relevant at weak coupling. The development of instabilities can be observed transparently in a renormalization group (RG) analysis \cite{isobe18prx,sherkunov18prb,lin19prb,hsu20prb}. With the Fermi surface nesting, the spin density waves can develop first at moderate RG scale, thereby trigger the true instabilities as the RG flow goes further \cite{lin19prb}. The leading instability usually occurs in the superconducting channels beyond $s$-wave. Meanwhile, the antiferromagnetic orders may become relevant at moderate coupling. However, recent experiments have uncovered correlated phases where the orders are more likely polarized \cite{sharpe19s,lu19n,serlin20s} or nematic \cite{kerelsky19n,jiang19n,cao20ax}. While the polarized orders may arise at moderate coupling as the conventional Stoner instability \cite{sherkunov18prb,hsu20prb,chichinadze20prb2} or through other mechanisms \cite{kang19prl,xie20prl,bultinck20prl,zhang19prr,wu20prl}, a theory where these zero-momentum density orders can develop as robust weak-coupling instabilities remains to be uncovered.

A potential answer to such problem is indicated by the emergence of `high-order' Van Hove singularity under particular setting, such as the magic angle \cite{yuan19nc}. In the moir\'e flat bands of twisted bilayer graphene, the number of saddle points can be different at different tunable parameters \cite{kim16nl,koshino18prx}. The variation of saddle point number amounts to the splitting of each saddle point into a pair. At the critical point of splitting, the saddle point becomes high-order, with the density of states acquiring a much stronger power-law divergence. Such divergence can lead to significantly different phase diagram from the one at the conventional Van Hove singularity. The manifestations of high-order Van Hove singularity has been investigated in the context of cuprate materials \cite{gofron94prl,isobe19prr}, doped intercalated graphene \cite{mcchesney10prl,gonzalez13prb,classen20prb}, strontium materials \cite{efremov19prl}, bilayer graphene \cite{shtyk17prb}, and twisted bilayer transition metal dichalcogenide \cite{bi19ax}. General discussions of the eligible band structures and locations of high-order Van Hove singularity have been conducted with complete classifications \cite{yuan20prb,chandrasekaran20prr}. For a single high-order Van Hove point, the renormalization group analysis uncovers an interacting fixed point \cite{isobe19prr}. This fixed point possesses divergent susceptibilities without developing long-range orders in various channels, thereby manifesting itself as a supermetal. When the high-order Van Hove singularity occurs at multiple points, the system may develop ferromagnetism as a leading weak-coupling instability of our interest \cite{shtyk17prb,classen20prb}. Superconductivity may also arise as a competing order. The indications from these works strongly suggest an analysis of high-order Van Hove singularity in twisted bilayer graphene, where polarized correlated phases may be uncovered.

In this work, we analyze the weakly repulsive electrons at the high-order Van Hove singularity in twisted bilayer graphene. Our study adopts the parquet RG analysis, which has been conducted at the conventional Van Hove singularity in square lattice \cite{schulz87el,dzyaloshinskii87jetp,furukawa98prl}, doped graphene \cite{nandkishore12np}, and twisted bilayer graphene \cite{isobe18prx,sherkunov18prb,lin19prb,hsu20prb}, as well as at the high-order Van Hove singularity in doped intercalated graphene \cite{classen20prb} and bilayer graphene \cite{shtyk17prb}. Unlike Refs.~\onlinecite{classen20prb, shtyk17prb} we consider a setting (relevant for twisted bilayer graphene) when the high-order Van Hove points occur away from the Brillouin zone boundary, which qualitatively alters the analysis. We show that such a system is primarily dominated by the zero-momentum particle-hole and particle-particle susceptibilities \cite{classen20prb}, leading to the RG flows toward either spin and/or valley-polarized orders or superconductivity. Five stable fixed trajectories under RG are uncovered, where the leading instabilities are $s$-wave ferromagnetism, $p$- and $d$-wave superconductivities, as well as $f$- and $p$-wave valley-polarized orders (Fig.~\ref{fig:pdwive}). We further examine the degeneracy breakdown in the multi-component irreducible pairing channels. The analysis shows that the chiral and helical orders are the energetically favored ground states in the $p$-wave superconductivity. Similarly, the chiral order is dominant in the $d$-wave superconductivity. Meanwhile, the polar order with spontaneous rotation symmetry breaking is favored in the $p$-wave valley-polarized order. The irrelevance of Fermi surface nesting suggest the stability of our results against band deformations preserving the high-order Van Hove singularity. Such feature is significantly different from the conventional Van Hove singularity, where the results may be fragile against the reduction of Fermi surface nesting. 

\section{High-order Van Hove singularity in twisted bilayer graphene}

\begin{figure}[t]
\centering
\includegraphics[scale = 1.02]{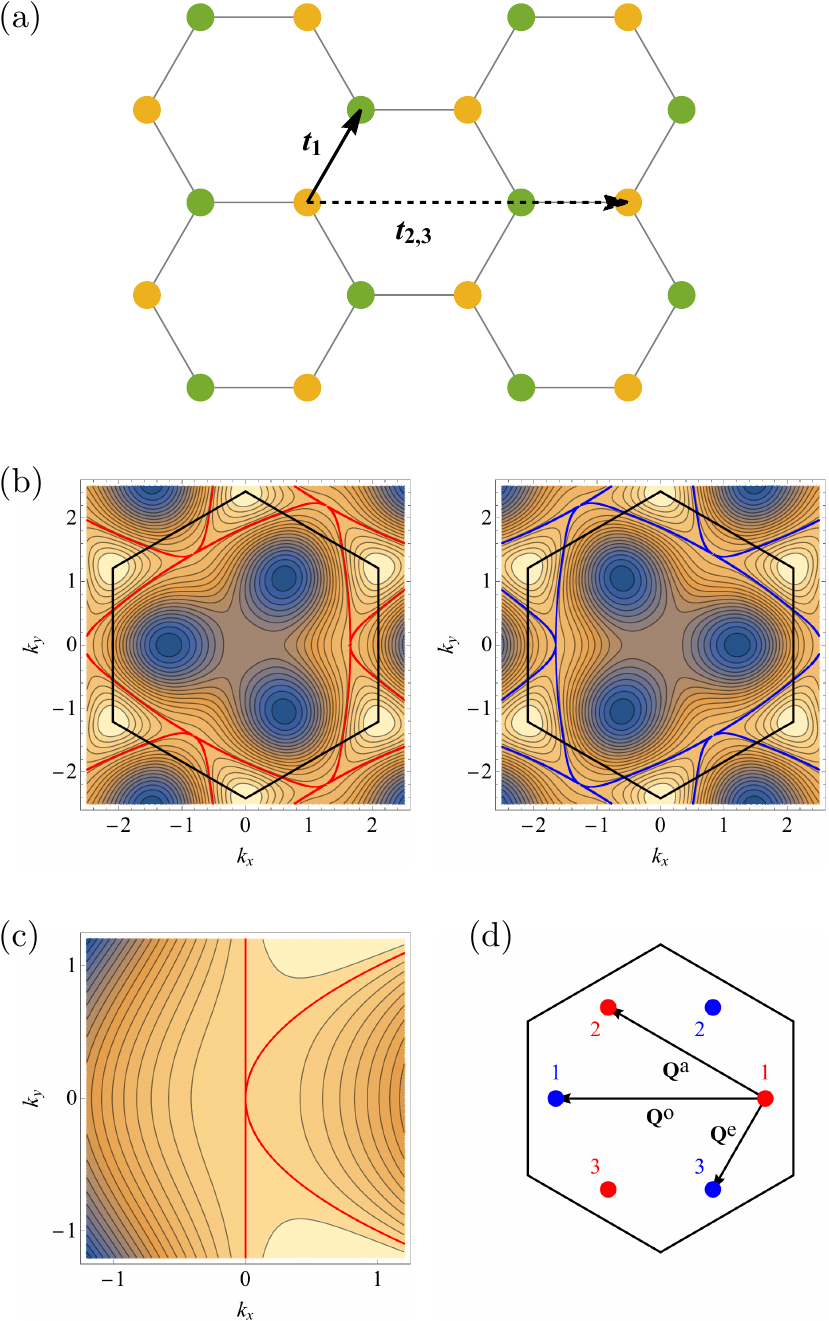}
\caption{\label{fig:tbg} High-order Van Hove singularity in twisted bilayer graphene and patch model. (a) Tight-binding honeycomb superlattice model for the nearly flat bands in twisted bilayer graphene \cite{yuan18prb}. The model contains the nearest- and the fifth-nearest-neighbor hoppings $t_{1,2}\sum_{\epvl{ij}_{1,5},\tau=\pm}(c_{i\tau}^\dag c_{j\tau}+\txt{H.c.})$, as well as an imaginary fifth-nearest-neighbor hopping $-it_3\sum_{\epvl{ij}_5,\tau=\pm}\tau c_{i\tau}^\dag c_{j\tau}+\txt{H.c.}$. Here $c_{i\tau}$ denotes the fermion operator at the site $i$ in the $\tau=\pm$ valley. (b) Band structure with high-order Van Hove singularity at $t_1=1$, $t_2=0.15$, and $t_3=0.2720717$. Each figure illustrates the band structure from a valley. Note that the high-order Van Hove singularity occurs on a `critical surface' in the three-dimensional phase spase spanned by $t_1$, $t_2$, and $t_3$. (c) The band structure (\ref{eq:dispersion}) in the vicinity of a high-order saddle point $\mbf P$ with $\k=0$. (d) The setup of patch model, where the patches are set at the high-order saddle points.}
\end{figure}

The model we consider is inspired by twisted bilayer graphene. In the twisted bilayer graphene at small twist angle, the low-energy regime is dominated by two pairs of conduction and valence moir\'e flat bands \cite{bistritzer11pnas,kim16nl,yuan18prb,koshino18prx,kang18prx,po18prx}. These moir\'e flat bands are manifest in the small moir\'e Brillouin zone, which corresponds to the large moir\'e superlattice in real space. Each pair of flat bands originates from the interlayer hybridization of Dirac cones in one graphene valley. The effective low-energy theory is described by a two-orbital honeycomb lattice model at moir\'e lattice scale [Fig.~\ref{fig:tbg}(a)], where the orbitals $\tau=\pm$ label the moir\'e flat bands from the two graphene valleys. The dispersion energies in the two valleys $\ve_{\pm,\mbf k}$ [Fig.~\ref{fig:tbg}(b)] are related under time-reversal symmetry $\ve_{-,-\mbf k}=\ve_{+,\mbf k}$, and an intravalley $\txt{C}_{3z}$ rotation symmetry is also manifest. The system obeys a spin $\txt{SO}(4)\sim\txt{SU}(2)_+\times\txt{SU}(2)_-$ symmetry composed of the spin $\txt{SU}(2)_\pm$ symmetries in the two valleys.

Van Hove singularity is generically present in the moir\'e flat bands of twisted bilayer graphene \cite{cao18n1,kerelsky19n,choi19np,xie19n,jiang19n}. We focus on the special case of `high-order' Van Hove singularity \cite{yuan19nc}, where the saddle points are at the critical point of splitting. Note that the critical point may manifest a subspace in the phase space of model parameters, where the band deformation does not break the high-order Van Hove singularity. Six high-order saddle points are present in the moir\'e Brillouin zone [Fig.~\ref{fig:tbg}(b)]. There are three points in each valley, sitting on the $\G M$ lines and exhibiting the dispersion energies [Fig.~\ref{fig:tbg}(c)]
\beeq
\label{eq:dispersion}
\ve_{\mbf P,\mbf k}=-\a k_\prll^2+\g k_\prll k_\perp^2+\k k_\perp^4.
\eneq
Here $k_\prll$ and $k_\perp$ denote the momentum deviations from each high-order saddle point $\mbf P$ parallel and perpendicular to the $\G M$ line, respectively. We have set the Van Hove doping at zero chemical potential $\mu=0$ for convenience. The points in the two valleys are related by time-reversal symmetry $\ve_{-\mbf P,-\mbf k}=\ve_{\mbf P,\mbf k}$. Note that the structure of high-order saddle point can be more easily seen in the representation $k_-=k_\prll-(\g/2\a)k_\perp^2$ and $k_+=k_\perp$
\beeq
\label{eq:hosdp}
\ve_{\mbf P,\mbf k}=A_+k_+^4-A_-k_-^2,
\eneq
where $A_+=\k+\g^2/4\a$ and $A_-=\a$. The dispersion energy is quadratic along one direction and quartic along the other one, with the sign changing for four times around the high-order saddle point $\mbf P$. The Fermi surface is determined by two parabolic curves $k_\prll=[(\g/2\a)\pm(\k/\a+\g^2/4\a^2)^{1/2}]k_\perp^2$, one of which becomes a straight line when $\k=0$. These curves touch with each other tangentially at the high-order saddle point $\mbf P$.

The high-order saddle points $\mbf P$'s manifest the high-order Van Hove singularity, where power-law divergence occurs in the density of states (Appendix \ref{app:hovhs}) \cite{yuan19nc,isobe19prr}
\beeq
D(\ve)\apx D_0\lf[\t(\ve)+\fr{1}{\sqrt2}\t(-\ve)\ri]|\ve|^{-1/4}.
\eneq
Here the prefactor is $D_0=\G(1/4)^2/(8\pi^{5/2}A_-^{1/2}A_+^{1/4})$. The power-law divergence is stronger than the logarithmic divergence at the conventional Van Hove singularity. Moreover, an asymmetry between the two sides of Van Hove doping can be observed, which is absent at the conventional Van Hove singularity. This feature originates from the different powers of momentum in the dispersion energies (\ref{eq:hosdp}) above and below the Van Hove doping.

Due to the power-law divergence in the density of states, the high-order saddle points $\mbf P$'s are more dominant than the other parts of the Fermi surface at low energy. We thus construct the low-energy theory by approximating the Fermi surface with six patches in the vicinity of these points. Such a ‘patch model’ takes the form \cite{isobe18prx,lin19prb}
\beeq
H=\sum_{\a\tau}(\ve_{\a\tau}-\mu)\psi_{\a\tau}^\dag\psi_{\a\tau}
\eneq
with the patch labels $\a=1,2,3$ [Fig.~\ref{fig:tbg}(d)]. The size of each patch is defined by an ultraviolet (UV) energy cutoff $\L$. In the patch model, the set of relevant momenta includes those between various pairs of patches. We define $\mbf Q^o$ as the momentum transfer between opposite patches, while $\mbf Q^{a,e}$ lie between patches with different patch labels in the same and different valleys, respectively.

\begin{figure}[b]
\centering
\includegraphics[scale = 1]{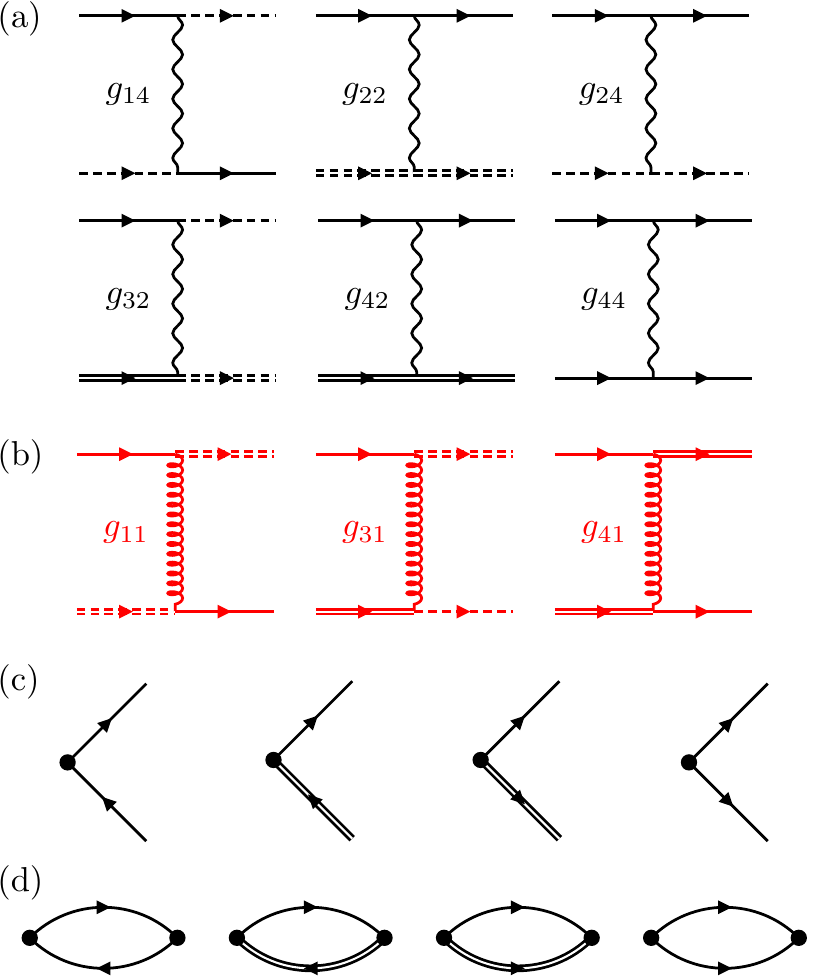}
\caption{\label{fig:fd} Feynman diagrams. (a) Six primary interactions without intervalley exchange. The solid and dashed lines describe the electrons from patches with different patch labels, while single and double lines characterize the electrons from different valleys. (b) The three interactions with intervalley exchange. (c) The test vertices in the particle-hole (first two) and particle-particle (last two) channels at momenta $\mbf0$ (first and third) and $\mbf Q^o$ (second and last). (d) Susceptibilities captured by the test vertices.}
\end{figure}

Assume that the interactions in the low-energy theory are weakly repulsive and spin $\txt{SU}(2)$ symmetric. It has been proposed that the interactions may be nonlocal on the moir\'e superlattice, thereby leading to attractions in some scattering channels \cite{kang19prl,chichinadze20prb1,chichinadze20prb2}. Projected on the patch model, the interactions can be classified into sixteen inequivalent types with different scattering processes among patches. These interactions can be labeled as $g_{ij}$
\beeq
H_\txt{int}=\fr{1}{2}\sum_{i,j=1}^4\sum_{\sst{\a_{1i},\a_{2i}\\\a_{3i},\a_{4i}}}\sum_{\sst{\tau_{1j},\tau_{2j}\\\tau_{3j},\tau_{4j}}}g_{ij}\psi_{\a_{1i}\tau_{1j}}^\dag\psi_{\a_{2i}\tau_{2j}}^\dag\psi_{\a_{3i}\tau_{3j}}\psi_{\a_{4i}\tau_{4j}},
\eneq
where $i$ and $j$ label the exchange, density-density, pair-hopping, and forward-scattering processes in the patch and valley sectors, respectively. The order of spins is $\s$, $\s'$, $\s'$, $\s$ in the interactions. Only nine interactions are eligible under momentum conservation. The interactions without intervalley exchange, including $g_{14}$, $g_{22}$, $g_{24}$, $g_{32}$, $g_{42}$, and $g_{44}$, are the primary ones considered in our analysis [Fig.~\ref{fig:fd}(a)]. These primary interactions are assumed repulsive at the bare level \cite{chichinadze20prb1,chichinadze20prb2}. Meanwhile, the interactions involving intervalley exchange, including $g_{11}$, $g_{31}$, and $g_{41}$, are assumed perturbative as they exhibit large momentum transfer at atomic scale [Fig.~\ref{fig:fd}(b)]. With the nonlocality in the interactions, these perturbative intervalley exchange may be either repulsive or attractive. We will include these when necessary to lift degeneracies, but not otherwise.

The high-order Van Hove singularity can lead to the breakdown of perturbation theory in Fermi liquid. This is manifest in the divergence of various static susceptibilities in the particle-hole (ph) and particle-particle (pp) channels $\Pi^\txt{ph/pp}_{\mbf q}=\mp T\sum_n\intv{k}G_{\mbf k\o_n}G_{(\pm\mbf k+\mbf q)(\pm\o_n)}$. Here $G_{\mbf k\o_n}=[i\o_n-(\ve_{\mbf k}-\mu)]^{-1}$ is the free fermion propagator with fermionic Matsubara frequency $\o_n=(2n+1)\pi T$. The Matsubara frequency summation leads to
\beeq
\label{eq:pippph}
\Pi^\txt{ph/pp}_{\mbf q}=-\intv{k}\fr{n_F(\ve_{\pm\mbf k+\mbf q}-\mu)-n_F(\pm[\ve_{\mbf k}-\mu])}{(\ve_{\pm\mbf k+\mbf q}-\mu)-[\pm(\ve_{\mbf k}-\mu)]},
\eneq
where $n_F(z)=[\exp(z/T)+1]^{-1}$ is the Fermi function. We calculate the susceptibilities in the patch model and focus on the asymptotic limit $\mu,T\ll\L$. The singularity in the density of states dominates in this regime, thereby selects a set of relevant susceptibilities with leading power-law divergence.

\begin{figure}[b]
\centering
\includegraphics[scale = 0.47]{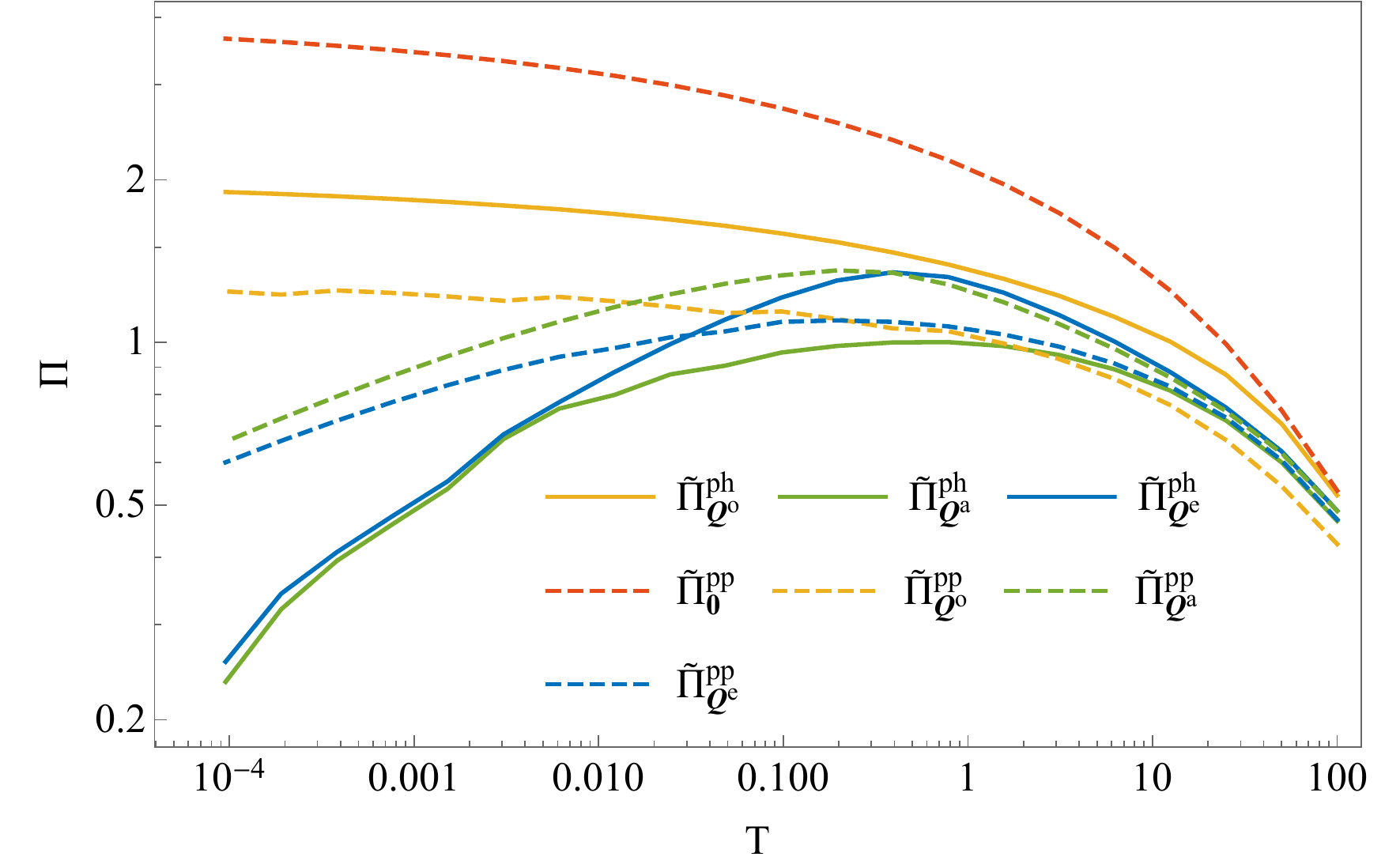}
\caption{\label{fig:pi} Particle-hole and particle-particle susceptibilities at the high-order Van Hove singularity. Here we assume maximal Fermi surface nesting at $\mbf q=\mbf Q^o$ by setting $\k=0$. Each curve indicates a rescaled static susceptibility $\til\Pi^\txt{ph/pp}_{\mbf q}=\Pi^\txt{ph/pp}_{\mbf q}/\Pi^0$.}
\end{figure}

Our analysis focuses on the Van Hove doping $\mu=0$. At zero momentum $\mbf q=\mbf0$, the particle-hole and particle-particle susceptibilities acquire the leading power-law divergences (Appendix \ref{app:hovhs})
\beeq
\label{eq:pi0}
\Pi^\txt{ph}_{\mbf 0}=\fr{1}{4}\Pi^\txt{pp}_{\mbf 0}\apx \Pi^0=D_0\fr{1+1/\sqrt2}{2}\fr{2.16514}{2}T^{-1/4}.
\eneq
The origins of these divergences are attributed to the singular density of states and Cooper divergence. We further compute all of the susceptibilities with decreasing temperature $T\rar0$ numerically (Fig.~\ref{fig:pi}). The results show that the leading power-law divergence also arises in both particle-hole and particle-particle channels at $\mbf q=\mbf Q^o$. Such divergence is expected from the observation of Fermi surface nesting, which becomes maximal when the Fermi surface contains a straight line at $\k=0$
\beeq
\Pi^\txt{ph}_{\mbf Q^o}\apx\fr{1}{2}\Pi^\txt{pp}_{\mbf 0},\quad
\Pi^\txt{pp}_{\mbf Q^o}\apx0.35\Pi^\txt{pp}_{\mbf 0}.
\eneq
An infrared (IR) cutoff is set by the finite $\k$ away from the maximal Fermi surface nesting. The rest of the susceptibilities are subleading and are irrelevant in the asymptotic limit.

We comment in passing on the situations when the doping is away from the high-order Van Hove singularity $\mu\neq0$. Since the Cooper divergence does not depend on the chemical potential, the according IR cutoff is always $T$. However, the divergences from singular density of states and Fermi surface nesting generically manifest the cutoff $\max\{\mu,T\}$. We thus expect $\Pi^\txt{pp}_{\mbf 0}$ as the most divergent susceptibility away from the Van Hove doping.

\section{Renormalization group}
\label{sec:rg}

With the divergent susceptibilities at the high-order Van Hove singularity, according divergence can occur in the interactions at low-energy. Such breakdown of perturbation theory in Fermi liquid can be captured by a parquet renormalization group (RG) analysis \cite{schulz87el,dzyaloshinskii87jetp,furukawa98prl,nandkishore12np,lin19prb}. The parquet RG analysis aims to identify the low-energy effective theory under renormalization. Starting from the UV cutoff $\L$ of the patch model, the shell of fast electron modes is progressively integrated out with decreasing temperature $T\rar0$. Such procedure leads to an evolving renormalized effective theory, where the interactions receive various one-loop corrections through the divergent susceptibilities. The evolution of the interactions form an RG flow toward a fixed point or a fixed trajectory. Making the standard `fast-parquet' approximation, we admit only the susceptibilities with the leading power-law divergence $\Pi^\txt{ph/pp}_{\mbf0\txt{/}\mbf Q^o}$. This approximation captures the RG flow in the asymptotic weak-coupling limit $g\rar0$.

The parquet RG procedure we employ is as follows. We assume the density of states diverges at a power $-\e$, which is treated as infinitesimal (but will ultimately be continued to $\epsilon = 1/4)$. At infinitesimal $\epsilon$, short-range interaction is marginal, and we therefore compute a set of RG equations to one-loop order. Note that such analysis is approximate and is an `$\e$-expansion' in the power of divergence \cite{classen20prb}. Define the dimensionless RG time $y=\ln(\L/T)$, interactions $\l_{ij}=\dot\Pi^\txt{pp}_{\mbf0}g_{ij}$, and relative susceptibilities $d^\txt{ph/pp}_{\mbf q}=\dot\Pi^\txt{ph/pp}_{\mbf q}/\dot\Pi^\txt{pp}_{\mbf0}$. While $d^\txt{ph}_{\mbf0}=0.25$ is intrinsic for the high-order Van Hove singularity herein, we assume $\k=0$ so that the maximal Fermi surface nesting at $\mbf q=\mbf Q^o$ leads to $d^\txt{ph}_{\mbf Q^o}=0.5$ and $d^\txt{pp}_{\mbf Q^o}=0.35$. Focusing on the primary interactions without intervalley exchange, we derive the RG equations with the form $\dot\l_{ij}=\b_{ij}(\{\l_{kl}\})$ (Appendix \ref{app:rg})
\beeq\beal
\label{eq:rgeqs}
\dot\l_{14}&=\e\l_{14}+d^\txt{ph}_{\mbf0}\l_{14}(\l_{14}+2\l_{44}),\\
\dot\l_{22}
&=\e\l_{22}+d^\txt{ph}_{\mbf0}[2\l_{22}(\l_{14}-2\l_{24}-\l_{44})\\&\hspace{11pt}+2\l_{42}(\l_{14}-2\l_{24})],\\
\dot\l_{24}
&=\e\l_{24}+d^\txt{ph}_{\mbf0}[2\l_{22}(-\l_{22}-2\l_{42})\\&\hspace{11pt}+2\l_{24}(\l_{14}-\l_{24}-\l_{44})+2\l_{14}\l_{44}],\\
\dot\l_{32}&=\e\l_{32}-\l_{32}(\l_{32}+2\l_{42}),\\
\dot\l_{42}
&=\e\l_{42}+d^\txt{ph}_{\mbf0}[4\l_{22}(\l_{14}-2\l_{24})-2\l_{42}\l_{44}]\\&\hspace{11pt}+d^\txt{ph}_{\mbf Q^o}\l_{42}^2-(2\l_{32}^2+\l_{42}^2),\\
\dot\l_{44}
&=\e\l_{44}+d^\txt{ph}_{\mbf0}[2\l_{14}(\l_{14}+2\l_{24})-4\l_{22}^2-4\l_{24}^2\\&\hspace{11pt}-2\l_{42}^2+\l_{44}^2]-d^\txt{pp}_{-\mbf Q^o}\l_{44}^2.
\enal\eneq
The first tree-level terms reflect the scaling dimension of the interactions, while the rest parts of the beta functions correspond to the one-loop corrections. With the setup of bare repulsion, the positive semi-definiteness constraint is imposed on $\l_{14}$ and $\l_{32}$ as indicated by their beta functions. The other interactions may flow in either positive or negative directions under RG.

We first examine the stability of the finite-coupling fixed points $\{\l_{ij}\}=\{\l_{ij}^*\}$ with $(\dot\l_{ij})_{\{\l_{kl}\}=\{\l_{kl}^*\}}=0$. In the vicinity of each fixed point, the linearized RG equations read $\d\dot\l_{ij}=M^\l_{ij,kl}\d\l_{kl}$ with $\d\l_{ij}=\l_{ij}-\l_{ij}^*$ and $M^\l_{ij,kl}=(\p\b_{ij}/\p\l_{kl})_{\{\l_{mn}\}=\{\l_{mn}^*\}}$. The eigenvalues of the matrix $M^\l=(M^\l_{ij,kl})$ indicate the flow directions along the eigenvectors. A negative eigenvalue indicates that the interactions flow toward the fixed point under RG, and vice versa. A stable fixed point is thus determined by the condition that all of the eigenvalues are negative. We find that all of the fixed points of the RG equations (\ref{eq:rgeqs}) are unstable. Our analysis thus focuses on the strong-coupling fixed trajectories, along which at least one of the interactions diverges at a finite scale $y_c$.

Along the fixed trajectories, the divergence of the interactions is captured by the critical scaling
\beeq
\label{eq:intcs}
\l_{ij}=\fr{\hat\l_{ij}}{y_c-y}.
\eneq
Adopting the critical scaling in the RG equations (\ref{eq:rgeqs}) leads to a set of algebraic equations for the critical interactions $\hat\l_{ij}$'s. These algebraic equations contain only the one-loop terms in the beta functions, since the tree-level terms become irrelevant along the strong-coupling fixed trajectories and vanish at $y=y_c$ in the algebraic equations. The potential fixed trajectories under RG can be identified with the solutions to these algebraic equations. Note that $\hat\l_{44}$ is finite for all of the nontrivial solutions. To examine the stability of the fixed trajectories, we analyze the RG flow of the reparametrized interactions $x_{ij}=\l_{ij}/\l_{44}$ for $ij\neq44$ with an alternative RG time $\l_{44}$ \cite{vafek10prb,nandkishore12np,lin19prb}
\beeq
\l_{44}\fr{dx_{ij}}{d\l_{44}}=\b^x_{ij}(\{x_{kl}\})=-x_{ij}+\fr{-\e x_{ij}+\b_{ij}(\{x_{kl}\})}{-\e x_{44}+\b_{44}(\{x_{kl}\})}.
\eneq
Here the tree-level terms in the orignal RG equations (\ref{eq:rgeqs}) are eliminated as they become irrelevant at divergent $\l_{44}$. The fixed points $\{x_{ij}\}=\{x_{ij}^*\}$ with $\l_{44}(dx_{ij}/d\l_{44})_{\{x_{kl}\}=\{x_{kl}^*\}}=0$ for these RG equations correspond to the fixed trajectories of the original RG equations (\ref{eq:rgeqs}). The stable fixed points are determined by having all of the eigenvalues of $M^x$ negative, where $M^x_{ij,kl}=(\p\b^x_{ij}/\p x_{kl})_{\{x_{mn}\}=\{x_{mn}^*\}}$. We find five different stable fixed trajectories compatible with bare respulsion. To which stable fixed trajectory the system flows under RG depends on the setup of bare interactions.

\section{Instability analysis}
\label{sec:inst}

The breakdown of perturbation theory at low-energy indicates that an instability to the Fermi liquid occurs. To probe the potential instabilities along the stable fixed trajectories, we introduce the test vertices in various particle-hole and particle-particle channels [Fig.~\ref{fig:fd}(c)] \cite{zanchi00prb,chubukov08prb,chubukov16prx,lin19prb}
\beeq
\label{eq:dh}
\d H=\sum[\D\psi^\dag\psi^{(\dag)}+\txt{H.c.}].
\eneq
The test vertices acquire corrections from the divergent susceptibilities under RG. Solving the flow equations of the test vertices (Appendix \ref{app:testvertex}), the irreducible pairing channels $I$'s are identified as the eigenmodes with
\beeq
\label{eq:tvflowipc}
\dot\D_I=-d_I\l_I\D_I.
\eneq
The interaction $\l_I$ in each channel is a linear combination of the interactions $\l_{ij}$'s in the patch model. Meanwhile, the susceptibility $d_I=d^\txt{ph/pp}_{\mbf q}$ is defined by the particle-hole/particle-particle type and the momentum $\mbf q$ of the channel. Along the stable fixed trajectories, the test vertices undergo the critical scaling $\D_I\sim(y_c-y)^{\b_I}$ as the interactions (\ref{eq:intcs}) do. The exponent in each channel is determined by the critical interaction and the susceptibility $\b_I=d_I\hat\l_I$.

\begin{table}[b]
\centering
\betb{|c|c|c|}
\hline
Channel & Pairings & Leading \\
\hline
$s$POM & $\psi^\dag\lf(\fr{\tau^0}{\sqrt2}\ri)\lf(\fr{\s^0}{\sqrt2}\ri)d_0\psi$ & \\
\hline
$f$VP & $\psi^\dag\lf(\fr{\tau^3}{\sqrt2}\ri)\lf(\fr{\s^0}{\sqrt2}\ri)d_0\psi$ & Yes\\
\hline
$d$POM & $\psi^\dag\lf(\fr{\tau^0}{\sqrt2}\ri)\lf(\fr{\s^0}{\sqrt2}\ri)d_{1,2}\psi$ & \\
\hline
$p$VP & $\psi^\dag\lf(\fr{\tau^3}{\sqrt2}\ri)\lf(\fr{\s^0}{\sqrt2}\ri)d_{1,2}\psi$ & Yes\\
\hline
$s$FM & $\psi^\dag\lf(\fr{\tau^0}{\sqrt2}\ri)\lf(\fr{\bsb\s}{\sqrt2}\ri)d_0\psi$ & Yes\\
\hline
$f$SVP & $\psi^\dag\lf(\fr{\tau^3}{\sqrt2}\ri)\lf(\fr{\bsb\s}{\sqrt2}\ri)d_0\psi$ & \\
\hline
$d$FM & $\psi^\dag\lf(\fr{\tau^0}{\sqrt2}\ri)\lf(\fr{\bsb\s}{\sqrt2}\ri)d_{1,2}\psi$ & \\
\hline
$p$SVP & $\psi^\dag\lf(\fr{\tau^3}{\sqrt2}\ri)\lf(\fr{\bsb\s}{\sqrt2}\ri)d_{1,2}\psi$ & \\
\hline
CDW$^o$ & $\psi^\dag\lf(\fr{\tau^{1,2}}{\sqrt2}\ri)\lf(\fr{\s^0}{\sqrt2}\ri)\psi$ & \\
\hline
SDW$^o$ & $\psi^\dag\lf(\fr{\tau^{1,2}}{\sqrt2}\ri)\lf(\fr{\bsb\s}{\sqrt2}\ri)\psi$ & \\
\hline
$s$SC & $\psi^\dag\lf(\fr{\tau^3}{\sqrt2}\ri)\lf(\fr{\s^0}{\sqrt2}\ri)d_0[i(i\tau^2)(i\s^2)(\psi^\dag)^T]$ & \\
\hline
$f$SC & $\psi^\dag\lf(\fr{\tau^0}{\sqrt2}\ri)\lf(\fr{\bsb\s}{\sqrt2}\ri)d_0[i(i\tau^2)(i\s^2)(\psi^\dag)^T]$ & \\
\hline
$d$SC & $\psi^\dag\lf(\fr{\tau^3}{\sqrt2}\ri)\lf(\fr{\s^0}{\sqrt2}\ri)d_{1,2}[i(i\tau^2)(i\s^2)(\psi^\dag)^T]$ & Yes \\
\hline
$p$SC & $\psi^\dag\lf(\fr{\tau^0}{\sqrt2}\ri)\lf(\fr{\bsb\s}{\sqrt2}\ri)d_{1,2}[i(i\tau^2)(i\s^2)(\psi^\dag)^T]$ & Yes\\
\hline
PDW$^o$ & $\psi^\dag\lf(\fr{\tau^{1,2}}{\sqrt2}\ri)\lf(\fr{\s^0}{\sqrt2}\ri)[i(i\tau^2)(i\s^2)(\psi^\dag)^T]$ & \\
\hline
\entb
\caption{\label{tb:tv} The irreducible pairing channels receiving leading power-law divergence and the particle-hole and particle-particle pairings therein. Here $\tau^\nu$ and $\s^\nu$ with $\nu=0,1,2,3$ are the Pauli matrices in the valley and spin pairing representations, respectively. The last column indicates whether the leading instability can develop in each channel starting from weakly repulsive primary interactions. Note that the effects of repulsive and attractive intervalley exchange have been considered.}
\end{table}

Our analysis focuses on the irreducible pairing channels which can receive the leading power-law divergence. These include the particle-hole and particle-particle channels at momenta $\mbf0$ and $\mbf Q^o$ (Table~\ref{tb:tv})
\beeq
\label{eq:instint}
\beal
\l_{s\txt{POM}\txt{/}f\txt{VP}}&=-2\l_{14}\pm4\l_{22}+4\l_{24}\pm2\l_{42}+\l_{44},\\
\l_{d\txt{POM}\txt{/}p\txt{VP}}&=\l_{14}\mp2\l_{22}-2\l_{24}\pm2\l_{42}+\l_{44},\\
\l_{s\txt{FM}\txt{/}f\txt{SVP}}&=-2\l_{14}-\l_{44},\\
\l_{d\txt{FM}\txt{/}p\txt{SVP}}&=\l_{14}-\l_{44},\\
\l_\txt{C/SDW$^o$}&=-\l_{42},\\
\l_\txt{$s$/$f$SC}&=2\l_{32}+\l_{42},\\
\l_\txt{$d$/$p$SC}&=-\l_{32}+\l_{42},\\
\l_\txt{PDW$^o$}&=\l_{44}.
\enal
\eneq
In the particle-hole branch, we have zero-momentum $s$- and $d$-wave Pomeranchuk orders ($s$/$d$POM), $f$- and $p$-wave valley-polarized orders ($f$/$p$VP), $s$- and $d$-wave ferromagnetisms ($s$/$d$FM), and $f$- and $p$-wave spin-valley-polarized orders ($f$/$p$SVP). The even- and odd-parity channels carry the valley singlet and triplet pairings $\tau^{0,3}$, respectively. The momentum-space form factors manifest the three irreducible patch representations under $\txt{C}_{3z}$ symmetry, including the nondegenerate $d_0=(1/\sqrt3)(1,1,1)$ and degenerate $d_1=(1/\sqrt6)(2,-1,-1)$, $d_2=(1/\sqrt2)(0,1,-1)$. There are also charge and spin density waves at $\mbf Q^o$ (C/SDW$^o$), where the valley triplet pairings $\tau^{1,2}$ are manifest and three degenerate orders can occur at the three momenta $\mbf Q^o$'s. On the other hand, the particle-particle branch contains $s$-, $f$-, $d$-, and $p$-wave superconductivities ($s$/$f$/$d$/$p$SC) at zero momentum. The even- and odd-parity channels now correspond to the valley triplet and singlet pairings $\tau^{3,0}$, respectively, and the irreducible patch representations $d_a$'s are again manifest. At the three $\mbf Q^o$'s, there are pair density waves (PDW$^o$) with valley triplet pairings $\tau^{1,2}$.

\begin{figure*}[t]
\centering
\includegraphics[scale = 1]{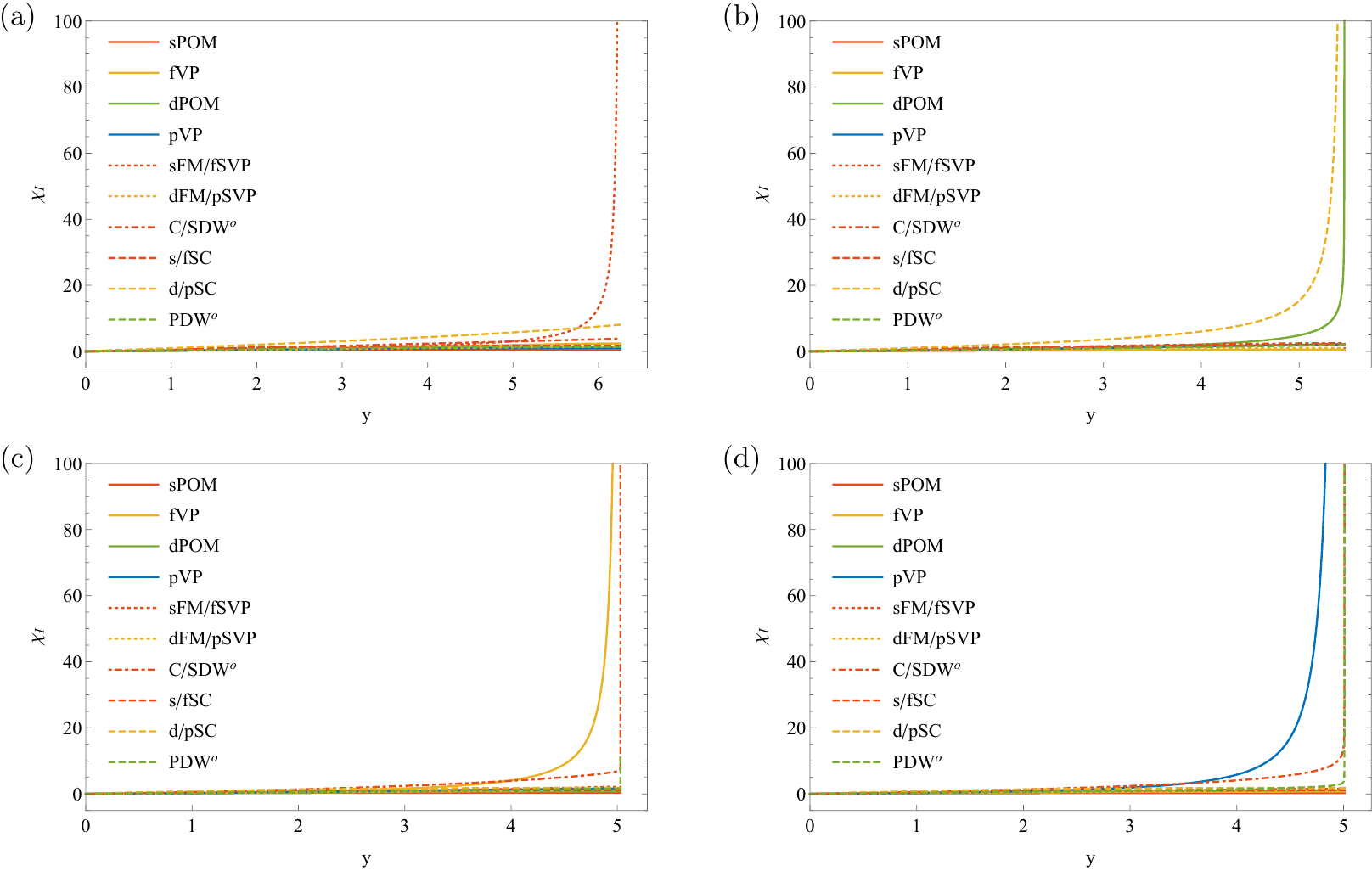}
\caption{\label{fig:rg} The susceptibilities of potential instabilities flow under RG. The maximal divergence occurs in the leading irreducible pairing channel at the critical scale $y=y_c$. Each figure illustrates the RG flow toward a stable fixed trajectory starting from weak repulsion, with (a) degenerate $s$-wave ferromagnetism/$f$-wave spin-valley-polarized order, (b) degenerate $d$/$p$-wave superconductivities, (c) $f$-wave valley-polarized order, or (d) $p$-wave valley-polarized order as the leading instability. The bare repulsions are set as $\l_{ij}=0.1$, except for (b) $\l_{24}=0.3$, (c) $\l_{42}=0.3$, and (d) $\l_{24}=\l_{42}=0.3$.}
\end{figure*}

To examine whether the instability occurs in a irreducible pairing channel, we probe the susceptibility with the test vertex in this channel [Fig.~\ref{fig:fd}(d)]
\beeq
\chi_I(y)=T\lf.\fr{\d^2\ln Z(y)}{\d\D_I(0)\d\bar\D_I(0)}\ri|_{\bar\D_I(0),\D_I(0),\psi^\dag,\psi=0}.
\eneq
Here $Z(y)$ denotes the partition function at scale $y$, which is obtained by integrating out the fast modes of electrons at $y'<y$. The overbar denotes the complex conjugate for the test vertex. The susceptibility undergoes the flows under RG
\beeq
\dot\chi_I=d_I\lf|\fr{\D_I}{\D_I(0)}\ri|^2
\eneq
and manifests the critical scaling $\chi_I\sim d_I(y_c-y)^{\a_I}$ along the stable fixed trajectories, as well. Here the exponent is determined by the test vertex exponent $\a_I=2\b_I+1$. The susceptibility becomes divergent at $y=y_c$ as $\a_I<0$, indicating the development of an instability. The leading instability occurs in the channel with the most divergent susceptibility, which manifests the most negative exponent $\a_I$ among all channels.

\begin{table}[b]
\centering
\betb{|c|c|c|c|c|}
\hline
Channel & $s$FM/$f$SVP & $d$/$p$SC & $f$VP & $p$VP \\
\hline
$\hat\l_{14}$ & 1.39963 & 0 & 0 & 0 \\
\hline
$\hat\l_{22}$ & 0 & 0 & 0.41863 & -0.60633 \\
\hline
$\hat\l_{24}$ & 0.69982 & 0 & -0.41863 & 0.60633 \\
\hline
$\hat\l_{32}$ & 0 & 0.42760 & 0 & 0 \\
\hline
$\hat\l_{42}$ & 0 & -0.71380 & 0.36359 & 0.93083 \\
\hline
$\hat\l_{44}$ & 1.30018 & -0.26160 & -0.43557 & -1.35101 \\
\hline
\entb
\caption{\label{tb:criticalint} The critical interactions along the four stable fixed trajectories starting from bare repulsion.}
\end{table}

\begin{figure}[t]
\centering
\includegraphics[scale = 1.02]{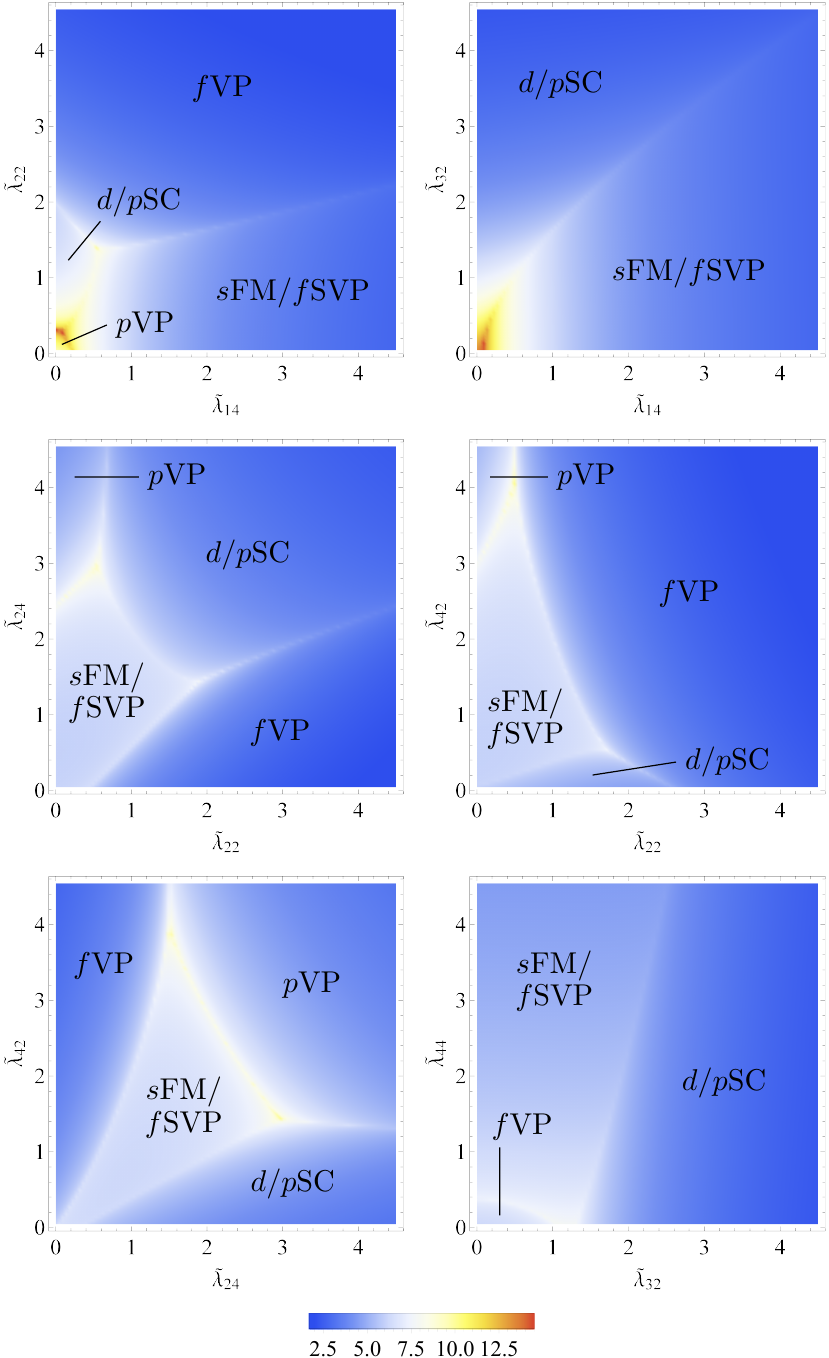}
\caption{\label{fig:pd} Phase diagram of the potential instabilities. The color map indicates the critical scale $y_c=\ln(\L/T_c)$, which is smaller in the phases and larger along the phase boundaries. In each two-interaction phase diagram, two of the interactions $\til\l_{ij}=\l_{ij}/\l_0$ are varied, and the rest ones are set at constant repulsion $\l_{ij}=\l_0=0.1$. The full set of two-interaction phase diagrams is demonstrated in Appendix \ref{app:pd}.}
\end{figure}

We identify the leading instabilities along the five stable fixed trajectories in our problem. These include degenerate $s$-wave ferromagnetism/$f$-wave spin-valley-polarized order, degenerate $d$/$p$-wave superconductivities, $f$-wave valley-polarized order, $p$-wave valley-polarized order, and $s$-wave Pomeranchuk order. We confirm that the first four instabilities are indeed accessible starting from bare repulsion (Fig.~\ref{fig:rg} and Table~\ref{tb:criticalint}). The phase diagram is further obtained under various setup of bare interactions (Fig.~\ref{fig:pd} and Appendix \ref{app:pd}). The interactions $\l_{14}$, $\l_{44}$ generically stabilize the degenerate $s$-wave ferromagnetism/$f$-wave spin-valley-polarized order. Meanwhile, increasing $\l_{22}$, $\l_{42}$ triggers the $f$-wave valley-polarized order. On the other hand, enlarged $\l_{24}$, $\l_{32}$ leads to the development of degenerate $d$/$p$-wave superconductivities. Finally, the $p$-wave valley-polarized order can occur in some regimes of the phase diagram. We do not see the $s$-wave Pomeranchuk order starting from bare repulsion, which may only be accessible when the bare attractions are involved.

It is interesting to discuss how the $d$/$p$-wave superconductivity arises from the bare repulsion. Although it is difficult to identify the pairing mechanism from the RG equations directly, we may acquire some intuition from the RG flow. Notably, we observe that the $d$-wave Pomeranchuk order also acquires a rapidly growing susceptibility in the $d$/$p$-wave superconducting phase. Near the phase boundary where the $d$/$p$-wave superconductivity just seizes the dominance, $\chi_{d\txt{POM}}$ even grows faster than $\chi_{d\txt{/}p\txt{SC}}$ before reaching the stable fixed trajectory under RG. Based on these observations, we interpret the formation of the $d$/$p$-wave superconductivity as driven by the $d$-wave Pomeranchuk order. While $d$-wave Pomeranchuk order grows first under RG, it shares the strength to the $d$/$p$-wave superconductivity gradually and turns it into the true instability along the stable fixed trajectory.

Importantly, the phase diagram is stable against the suppression of Fermi surface nesting by $\k\neq0$ in the dispersion energy (\ref{eq:dispersion}). This indicates that the key features are determined primarily by the zero-momentum particle-hole and particle-particle susceptibilities. Although the phase diagram may be altered by the finite-momentum particle-hole and particle-particle susceptibilities $\Pi^\txt{ph/pp}_{\mbf Q^o}$ at $d^\txt{ph}_{\mbf0}\neq0.25$ or higher Fermi surface nesting $d^\txt{ph}_{\mbf Q^o}>0.5$ and $d^\txt{pp}_{\mbf Q^o}>0.35$, these regimes are beyond the accessible range of physical systems in the asymptotic limit and are thus excluded from our analysis. With the stability against band deformations, our results may be robustly applicable across the twisted bilayer graphene systems under various conditions which contain the high-order Van Hove singularity.

\begin{figure*}[t]
\centering
\includegraphics[scale = 1]{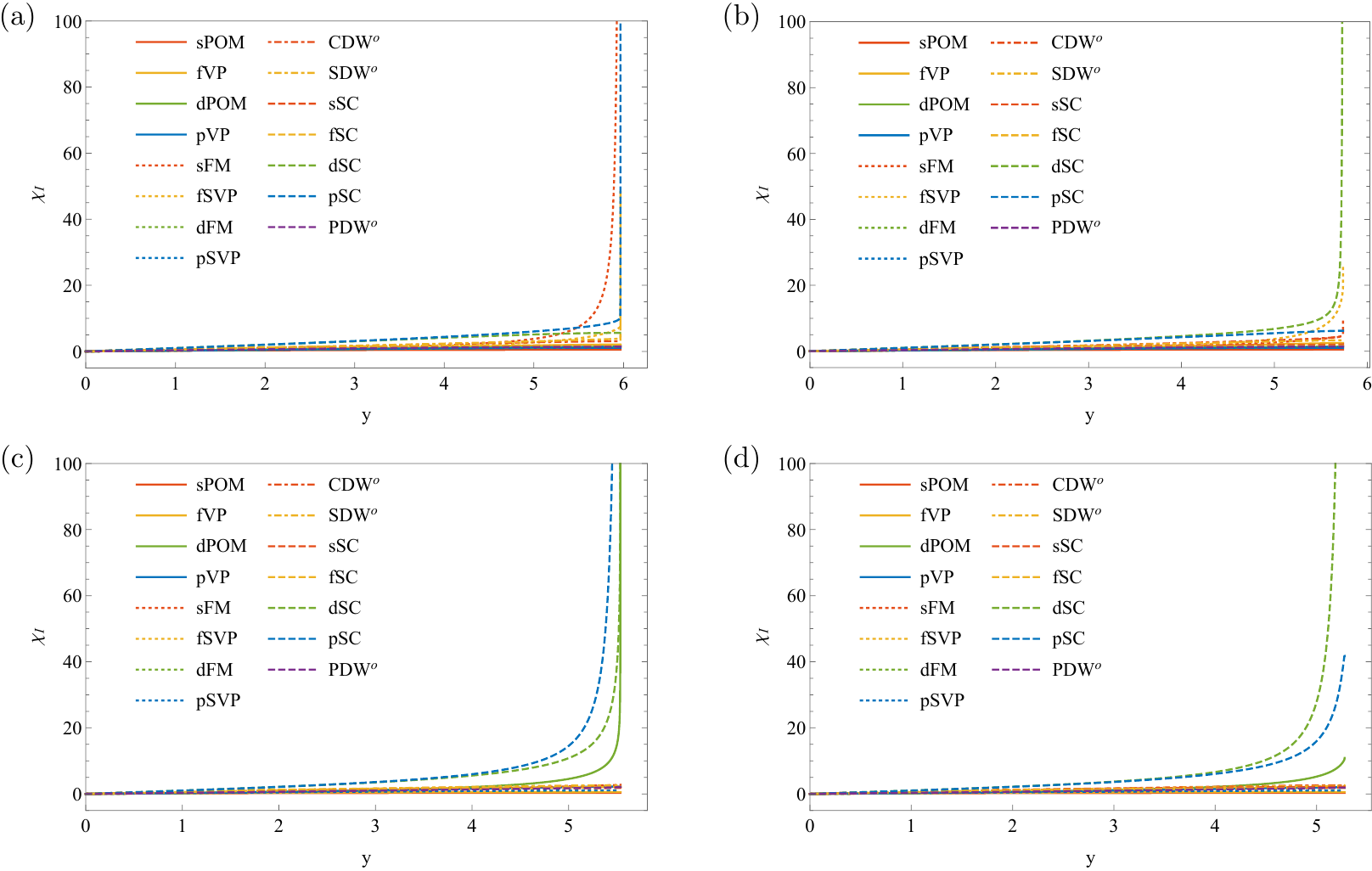}
\caption{\label{fig:rgwive} The breakdown of degeneracy between different instabilities by intervalley exchange. The bare values of the primary repulsions in (a) and (b) [(c) and (d)] are set the same as in Fig~\ref{fig:rg}(a) [Fig~\ref{fig:rg}(b)]. Meanwhile, the intervalley exchange is taken as perturbative repulsion (attraction) $\l_{i1}=(-)0.01$ in (a) and (c) [(b) and (d)]. With repulsive intervalley exchange, (a) the $s$-wave ferromagnetism wins over the $f$-wave spin-valley-polarized order, while (c) the $p$-wave superconductivity beats the $d$-wave one. On the other hand, the $d$-wave superconductivity beats (b) the $f$-wave spin-valley-polarized order and (d) the $p$-wave superconductivity when the interalley exchange is attractive.}
\end{figure*}

The phase diagram can be contrasted with the one of an $\txt{SU}(4)$ symmetric model without valley splitting \cite{classen20prb}. In such model, the high-order saddle points sit at the centers of Brillouin zone boundaries. The correspondence between the instabilities in the two models can be identified. The particle-hole instabilities with spin and/or valley polarized orders all correspond to the $\txt{SU}(4)$ flavor ferromagnetism. On the other hand, the even- and odd-parity pairings of superconductivities all correspond to the odd-exchange $\txt{SU}(4)$ flavor pairings. The zero-momentum irreducible pairing channels in the $\txt{SU}(4)$ symmetric model are classified primarily by the momentum-space form factors $d_a$'s, which are either $s$- or $d$-wave. With the dominance of zero-momentum particle-hole and particle-particle susceptibilities in both models, a similarity between the phase diagrams is expected. However, the valley splitting generically leads to the distinctions between some correlated phases in twisted bilayer graphene, which are absent in the $\txt{SU}(4)$ symmetric model.

\section{What intervalley exchange does}
\label{sec:wive}

\begin{figure}[t]
\centering
\includegraphics[scale = 1]{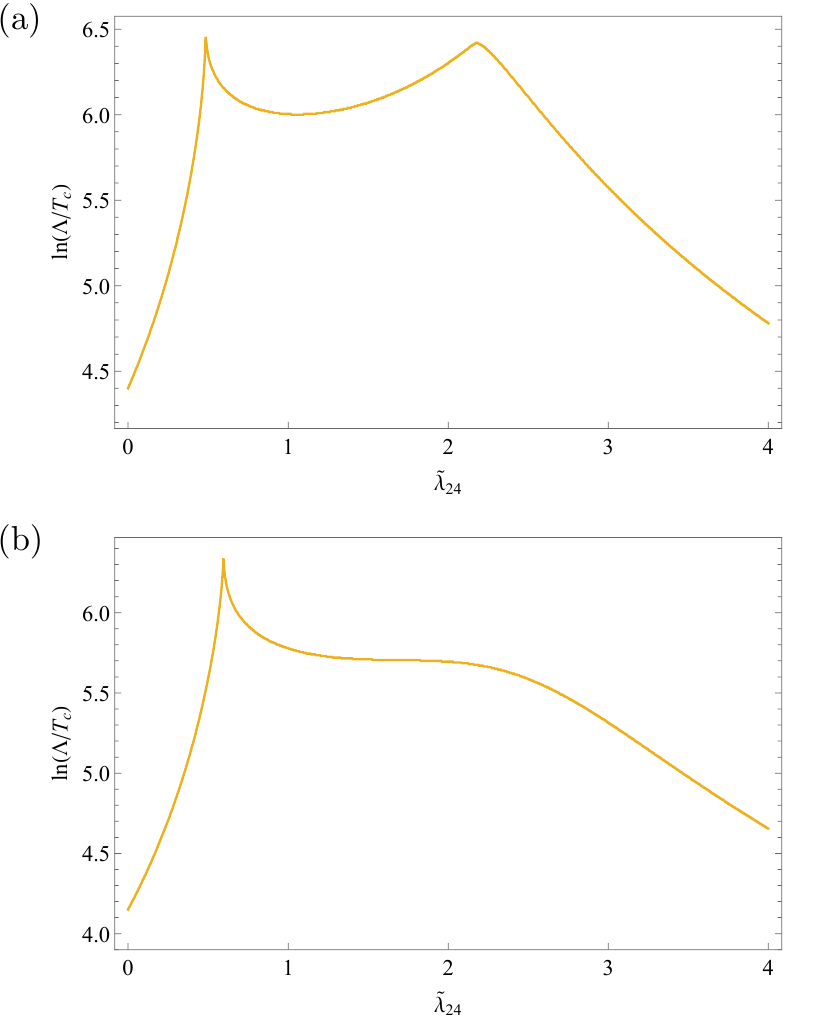}
\caption{\label{fig:ptwive} The phase distinction with (a) repulsive and (b) attractive intervalley exchange $\l_{i1}=\pm0.01$. Here $\til\l_{24}=\l_{24}/\l_0$ is varied, while the other primary interactions are constant $\l_{ij\neq24}=\l_0=0.1$. (a) With repulsive intervalley exchange, the $s$-wave ferromagnetism and the $p$-wave superconductivity form distinct phases separated by a peak of $y_c$. The three phases are (left) $f$-wave valley-polarized order, (center) $s$-wave ferromagnetism, and (right) $p$-wave superconductivity [Fig.~\ref{fig:pdwive}(a)]. (b) When the intervalley exchange is attractive, the right peak vanishes. This indicates that the $d$-wave superconductivity beats the $f$-wave spin-valley-polarized order and occupies the according phase regime [Fig.~\ref{fig:pdwive}(b)].}
\end{figure}

To address the more realistic situations in the practical twisted bilayer graphene systems, we now include the intervalley exchange perturbatively $|g_{i1}|\ll g_{i2}, g_{i4}$ [Fig.~\ref{fig:fd}(b)] and examine their effects on the correlated phases. This introduces additional terms to the RG equations (\ref{eq:rgeqs}) as well as to the interactions in the irreducible pairing channels (\ref{eq:instint}) (Appendices \ref{app:rg} and \ref{app:testvertex}). Since the intervalley exchange breaks the $\txt{SU}(2)_+\times\txt{SU}(2)_-$ symmetry down to $\txt{SU}(2)$, it can break the original degeneracies between some irreducible pairing channels \cite{isobe18prx,hsu20prb}. The competition between different instabilities may also change qualitatively under such perturbation.

When the intervalley exchange is repulsive $\l_{i1}>0$, the original phase diagram (Fig.~\ref{fig:pd}) does not experience qualitative change except for the breakdown of degeneracy. Along the stable fixed trajectory with degenerate polarized orders, $s$-wave ferromagnetism wins over the $f$-wave spin-valley-polarized order [Fig.~\ref{fig:rgwive}(a)]. Meanwhile, $p$-wave superconductivity beats the $d$-wave one along another stable fixed trajectory [Fig.~\ref{fig:rgwive}(c)]. Note that the valley singlet pairing is generically favored, as can be expected for the repulsive intervalley exchange. Things become different when the intervalley exchange is attractive. Along the stable fixed trajectory with degenerate superconductivity, the $d$-wave superconductivity now beats the $p$-wave one [Fig.~\ref{fig:rgwive}(d)]. This result is expected since the attractive intervalley exchange favors the valley triplet pairing. It is tempting to expect that the $f$-wave spin-valley-polarized order wins over the $s$-wave ferromagnetism and dominates another stable fixed trajectory. However, the fixed trajectory analysis (Table~\ref{tb:criticalintwIVE}) indicates that such stable fixed trajectory is absent. What occurs instead is that the $d$-wave superconductivity defeats the $f$-wave spin-valley-polarized order and occupies the according phase regime [Fig.~\ref{fig:rgwive}(b)]. We justify the expansion of superconducting regime by measuring the critical scale $y_c$ under RG. With repulsive intervalley exchange, the $s$-wave ferromagnetism is separated from the $p$-wave superconductivity by a peak [Fig.~\ref{fig:ptwive}(a)]. On the other hand, the two phase regimes are smoothly connected without a peak in-between under attractive intervalley exchange [Fig.~\ref{fig:ptwive}(b)]. This indicates that both phase regimes are occupied by the $d$-wave superconductivity. 

Summarizing these results, we arrive at a tentative electronic phase diagram (Fig.~\ref{fig:pdwive}) at the high-order Van Hove singularity in twisted bilayer graphene. Five correlated phases arise under repulsive primary interactions and repulsive/attractive intervalley exchange, including $s$-wave ferromagnetism, $p$- and $d$-wave superconductivities, as well as $f$- and $p$-wave valley-polarized orders.

\begin{table}[b]
\centering
\betb{|c|c|c|c|c|c|}
\hline
Channel & $s$FM & $p$SC & $d$SC & $f$VP & $p$VP \\
\hline
$\hat\l_{14}$ & 0.73516 & 0 & 0 & 0 & 0 \\
\hline
$\hat\l_{22}$ & 0.36758 & 0 & 0 & 0.41863 & -0.60633 \\
\hline
$\hat\l_{24}$ & 0.36758 & 0 & 0 & -0.41863 & 0.60633 \\
\hline
$\hat\l_{32}$ & 0 & 0.24073 & 0.16911 & 0 & 0 \\
\hline
$\hat\l_{42}$ & 0.16937 & -0.47477 & -0.29812 & 0.36359 & 0.93083 \\
\hline
$\hat\l_{44}$ & 0.81745 & -0.16074 & 0.04507 & -0.43557 & -1.35101 \\
\hline
$\hat\l_{11}$ & 0.73516 & 0 & 0 & 0 & 0 \\
\hline
$\hat\l_{31}$ & 0 & -0.24073 & 0.16911 & 0 & 0 \\
\hline
$\hat\l_{41}$ & 0.44739 & 0.26596 & -0.37100 & 0 & 0 \\
\hline
\entb
\caption{\label{tb:criticalintwIVE} The critical interactions along the five stable fixed trajectories in the presence of intervalley exchange.}
\end{table}

\section{The ordered states}
\label{sec:order}

Our RG analysis has uncovered five correlated phases that can develop at the high-order Van Hove singularity in twisted bilayer graphene. We now proceed to discuss the various interesting features that these phases can possess.

\subsection{Nondegenerate polarized orders}

For the $s$-wave ferromagnetism, the irreducible representation in the patch sector is $d_0$. The order parameter breaks the spin $\txt{SU}(2)$ symmetry spontaneously
\beeq
\mbf\D=\lf\la\psi^\dag\lf(\fr{\tau^0}{\sqrt2}\ri)\lf(\fr{\bsb\s}{\sqrt2}\ri)d_0\psi\ri\ra,
\eneq
which serves as a spontaneous Zeeman splitting
\beeq
\D_{\mbf k}=\fr{1}{2}\mbf\D\cdot\tau^0\bsb\s
\eneq
and separates the spin-up and down Fermi surfaces. The axial direction of spin lies along an arbitrary direction. Things are similar in the $f$-wave valley-polarized order, where the order parameter is
\beeq
\D=\lf\la\psi^\dag\lf(\fr{\tau^3}{\sqrt2}\ri)\lf(\fr{\s^0}{\sqrt2}\ri)d_0\psi\ri\ra.
\eneq
The only difference is that the valley polarization is now fixed instead of being arbitrary, and the gap function is
\beeq
\D_{\mbf k}=\fr{1}{2}\D\tau^3\s^0.
\eneq

\subsection{$p$-wave superconductivity}

The $p$-wave superconductivity manifests two degenerate complex vector order parameters
\beeq
\mbf\D_a^\dag=\lf\la\psi_+^\dag\lf(\fr{\bsb\s}{\sqrt2}\ri)d_a[i(i\s^2)(\psi_-^\dag)^T]\ri\ra
\eneq
with $a=1,2$, which may support more interesting features then the nondegenerate channels. Note that the irreducible valley pairing representation has been reduced, and $\psi_\pm$ denote the electrons in the two valleys. The Ginzburg-Landau free energy is derived in the vicinity of mean-field critical temperature $T_c$ (Appendix \ref{app:gl})
\beeq\label{eq:freeengpsc}\beal
f&=c^{(2)}(|\mbf\D_1|^2+|\mbf\D_2|^2)+c^{(4)}\bigg\{(|\mbf\D_1|^2+|\mbf\D_2|^2)^2\\&\quad\,+|\mbf{\bar\D}_1\times\mbf\D_1|^2+|\mbf{\bar\D}_2\times\mbf\D_2|^2\\&\quad\,+\fr{1}{3}[-2|\mbf\D_1|^2|\mbf\D_2|^2+\mbf\D_1^2\mbf{\bar\D}_2^2+\mbf{\bar\D}_1^2\mbf\D_2^2\\&\quad\,-2(\mbf\D_1\times\mbf{\bar\D}_2)^2-2(\mbf{\bar\D}_1\times\mbf\D_2)^2\\&\quad\,+4|\mbf{\bar\D}_1\times\mbf\D_2|^2-4|\mbf\D_1\times\mbf\D_2|^2]\bigg\}.
\enal\eneq
In accordance with the development of superconductivity below $T_c$, the quadratic prefactor $c^{(2)}\sim T-T_c$ turns negative while the quartic prefactor $c^{(4)}$ remains positive. The types of energetically favored ground states are determined by the anisotropic terms at quartic order. The second line penalizes the chiral spin orders, thereby confines the order parameters in the polar spin forms $\mbf\D_a=\D_a\mbf{\hat n}_a$ with axial unit vector $\mbf{\hat n}_a$. This further eliminates the last line since both terms exhibit $|\D_1|^2|\D_2|^2|\mbf{\hat n}_1\times\mbf{\hat n}_2|^2$. Define $\t=\cos^{-1}(\mbf{\hat n}_1\cdot\mbf{\hat n}_2)$ as the angle between the order parameters and $\phi=\txt{Arg}(\D_2/\D_1)$ as the phase difference. The free energy reduces to
\beeq\beal
f&=c^{(2)}(|\mbf\D_1|^2+|\mbf\D_2|^2)+c^{(4)}\bigg[(|\mbf\D_1|^2+|\mbf\D_2|^2)^2\\&\quad\,-\fr{4}{3}|\mbf\D_1|^2|\mbf\D_2|^2(\cos^2\t\sin^2\phi+\sin^2\t\cos^2\phi)\bigg],
\enal\eneq
from which the energetically favored ground states can be identified directly.

We see two ground states by inspecting the second line of the free energy. The first ground state is the $p\pm ip$ chiral order ($p$CSC) at $
\t=0,\pi$ and $\phi=\pm\pi/2$, where $\mbf{\hat n_1}=\pm\mbf{\hat n}_2=\mbf{\hat n}$ and $\D_2=\pm i\D_1$. This state breaks the time-reversal symmetry spontaneously and manifests the gap function
\beeq
\D^\txt{C}_\pm(\mbf{\hat n})=\fr{1}{2}\D\tau^0(d_1\pm id_2)(\mbf{\hat n}\cdot\bsb\s),
\eneq
whose phase winds for $\pm2\pi$ around the Fermi surface
\beeq
\D^\txt{C}_{\pm,\t_{\mbf k}}(\mbf{\hat n})=\fr{1}{\sqrt6}\D e^{\pm i\t_{\mbf k}}(\mbf{\hat n}\cdot\bsb\s)
\eneq
with the polar angle $\t_{\mbf k}$. The axial direction of the order parameter $\mbf{\hat n}$ can lie in an arbitrary direction, corresponding to the spontaneous breakdown of spin $\txt{SU}(2)$ symmetry. On the other hand, the second ground state is the $p$-wave helical order ($p$HSC) at $
\t=\pi/2$ and $\phi=0,\pi$, with $\mbf{\hat n_1}\cdot\mbf{\hat n}_2=0$ and $\D_1=\pm\D_2$. The time-reversal symmetry is preserved, and the gap function is
\beeq
\D^\txt{H}(\mbf{\hat n}_1,\mbf{\hat n}_2)=\fr{1}{2}\D\tau^0(d_1\mbf{\hat n}_1+d_2\mbf{\hat n}_2)\cdot\bsb\s
\eneq
with the momentum-space form
\beeq
\D^\txt{H}_{\t_{\mbf k}}(\mbf{\hat n}_1,\mbf{\hat n}_2)=\fr{1}{\sqrt6}\D(\cos\t_{\mbf k}\mbf{\hat n}_1+\sin\t_{\mbf k}\mbf{\hat n}_2)\cdot\bsb\s.
\eneq
This ground state can be regarded as a composition of two chiral orders, where the equal-spin pairings with opposite spins $\uar\uar,\dar\dar$ exhibit opposite phase windings $\pm2\pi$. The spin $\txt{SU}(2)$ symmetry is again broken spontaneously, and a $\pm2\pi$ winding in the plane formed by $\mbf{\hat n}_1$ and $\mbf{\hat n}_2$ can be observed around the Fermi surface.

The superconducting ground states exhibit fully gapped quasiparticle spectra and belong to distinct $\mbb Z$ topological classification \cite{xu18prl,you19npjqm}. Since the two ground states sit at disjoint free energy minima in the order parameter space, a first-order phase transition is expected to occur in between.

\subsection{$d$-wave superconductivity}

For the $d$-wave superconductivity, the two degenerate complex scalar order parameters
\beeq
\D_a^\dag=\lf\la\psi_+^\dag\lf(\fr{\s^0}{\sqrt2}\ri)d_a[i(i\s^2)(\psi_-^\dag)^T]\ri\ra
\eneq
with $a=1,2$ are manifest. Here the irreducible valley pairing representation is again reduced. The Ginzburg-Landau free energy in the vicinity of mean-field critical temperature $T_c$ reads (Appendix \ref{app:gl}) \cite{nandkishore12np,lin18prb}
\beeq\label{eq:freeengdsc}\beal
f&=c^{(2)}(|\D_1|^2+|\D_2|^2)+c^{(4)}\bigg[(|\D_1|^2+|\D_2|^2)^2\\&\quad\,+\fr{1}{3}(-2|\D_1|^2|\D_2|^2+\D_1^2\bar\D_2^2+\bar\D_1^2\D_2^2)\bigg].
\enal\eneq
In accordance with the development of superconductivity below $T_c$, the quadratic prefactor $c^{(2)}\sim T-T_c$ turns negative while the quartic prefactor $c^{(4)}$ remains positive. The types of energetically favored ground states are determined by the anisotropic terms at quartic order. With $\phi=\txt{Arg}(\D_2/\D_1)$ defined as the phase difference between the order parameters, the free energy reduces to
\beeq\beal
f&=c^{(2)}(|\D_1|^2+|\D_2|^2)+c^{(4)}\bigg[(|\D_1|^2+|\D_2|^2)^2\\&\quad\,-\fr{4}{3}|\D_1|^2|\D_2|^2\sin^2\phi\bigg].
\enal\eneq
We identify the ground state as the $d\pm id$ chiral order ($d$CSC) at $\phi=\pm\pi/2$, which indicates $\D_2=\pm i\D_1$. The gap function
\beeq
\D^\txt{C}_\pm=\fr{1}{2}\D\tau^3(d_1\pm id_2)\s^0
\eneq
manifests $\pm4\pi$ phase winding around the Fermi surface
\beeq
\D^\txt{C}_{\pm,\t_{\mbf k}}=\fr{1}{\sqrt6}\D e^{\pm i2\t_{\mbf k}}\s^0,
\eneq
indicating the spontaneous breakdown of time-reversal symmetry. This state exhibits fully gapped quasiparticle spectrum and belongs to $\mbb Z$ topological classification \cite{xu18prl,you19npjqm}. Note that the spin $\txt{SU}(2)$ symmetry is preserved by the spin singlet pairing.

\begin{table*}[t]
\centering
\betb{|c|c|c|c|}
\hline
Phase & Broken symmetry & Low-energy spectrum & Experimental probe \\
\hline
$s$FM & $\txt{SU}_s(2)$, $\mca T$ & Spin-splitted FS & Magnetic susceptibility\\
\hline
$p$CSC & $\txt{U}_c(1)$, $\txt{SU}_s(2)$, $\mca T$ & Fully gapped & Spin and thermal quantum Hall effects, polar Kerr effect\\
\hline
$p$HSC & $\txt{U}_c(1)$, $\txt{SU}_s(2)$ & Fully gapped & Nontrivial Josephson coupling\\
\hline
$d$CSC & $\txt{U}_c(1)$, $\mca T$ & Fully gapped & Spin and thermal quantum Hall effects, polar Kerr effect\\
\hline
$f$VP & $\txt{U}_v(1)$, $\mca T$ & Valley-splitted FS & Valley Hall effect\\
\hline
$p$PVP & $\txt{U}_v(1)$, $\mca T$, $C_{3z}$ & Anisotropic valley-splitted FS & Valley Hall effect, anisotropic LDOS or transport signal\\
\hline
\entb
\caption{\label{tb:phase} The potential correlated phases in our model and their features. Here $\mca T$ is the time-reversal symmetry, while the symmetries with subscripts $c$, $s$, and $v$ are those in the charge, spin, and valley sectors, respectively. FS denotes the Fermi surface, and LDOS refers to the local density of states.}
\end{table*}

\subsection{$p$-wave valley-polarized order}

The two degenerate real order parameters in the $p$-wave valley-polarized order are 
\beeq
\D_a=\lf\la\psi^\dag\lf(\fr{\tau^3}{\sqrt2}\ri)\lf(\fr{\s^0}{\sqrt2}\ri)d_a\psi\ri\ra
\eneq
with $a=1,2$. We derive the Ginzburg-Landau free energy in the vicinity of mean-field critical temperature $T_c$ (Appendix \ref{app:gl})
\beeq
\label{eq:freeengpvp}
\beal
f&=c^{(2)}(\D_1^2+\D_2^2)+c^{(4)}(\D_1^2+\D_2^2)^2\\&\hspace{11pt}+c^{(6)}(\D_1^2+\D_2^2)^3\lf(1+\fr{1}{10}\cos6\t_\D\ri)\\&\hspace{11pt}+c^{(8)}(\D_1^2+\D_2^2)^4\lf(1+\fr{8}{35}\cos6\t_\D\ri).
\enal
\eneq
Here the angle $\t_\D=\tan^{-1}(\D_2/\D_1)$ is defined. The quadratic prefactor $c^{(2)}\sim T-T_c$ turns negative below $T_c$, where the valley-polarized order develops. At higher orders in the expansion, the prefactors $c^{(4)}$ and $c^{(8)}$ are positive, while $c^{(6)}$ is negative. The anisotropic terms exist and select particular directions under spontaneous $\txt{C}_{3z}$ rotation symmetry breaking \cite{fernandes20sa}. Since the octic order terms are perturbatively smaller than the sextic order ones, the free energy minima occur at $\t_\D=n\pi/3$ with $n=0,1,2$. The gap function of such polar order ($p$PVP)
\beeq\beal
\D(\t_\D)=\fr{1}{2}\D\tau^3(d_1\cos\t_\D+d_2\sin\t_\D)\s^0
\enal\eneq
manifests one of the patch orders $(2,-1,-1)$, $(-1,2,-1)$, and $(-1,-1,2)$, and the momentum-space form reads
\beeq
\D_{\t_{\mbf k}}(\t_\D)=\fr{1}{\sqrt6}\D\cos(\t_{\mbf k}-\t_\D)\s^0.
\eneq
Accordingly, the Fermi surface undergoes a deformation which is anisotropic in the momentum space.

\section{Discussion}

We have analyzed the correlated phases that may arise as weak-coupling instabilities when multiple high-order Van Hove points occur within the Brillouin zone, in a model inspired by twisted bilayer graphene. The parquet renormalization group analysis uncovers five different correlated phases starting from weakly repulsive primary interactions and secondary intervalley exchange (Table \ref{tb:phase}). These include $s$-wave ferromagnetism, $p$-wave chiral/helical superconductivity, $d$-wave chiral superconductivity, $f$-wave valley-polarized order, and $p$-wave polar valley-polarized order. The Fermi surfaces are present in the spin- and valley-polarized orders with splittings and/or anisotropic deformations, while the chiral and helical superconductivities are fully gapped. Significantly, the phase diagram is determined primarily by the zero-momentum particle-hole and particle-particle susceptibilities. This indicates the stability of our results against band deformations which preserve the high-order Van Hove singularity. Our work thus serves as a potential guide toward the understanding of experimentally observed correlated phases in twisted bilayer graphene under various conditions.

It is worth discussing the experimental manifestations of the correlated phases we uncover. The spin- and valley-polarized orders may be observed from the measurements of magnetic susceptibility and valley Hall effect, respectively. When spatial rotation symmetry is broken, the according anisotropy can be observed in the probe of local density of states \cite{kerelsky19n,jiang19n} or transport measurement \cite{cao20ax}. The chiral and helical superconductivities can manifest topological responses. While the chiral ordered states exhibit the spin and thermal quantum Hall effects \cite{senthil99prb,horovitz03prb,sengupta06prb} and polar Kerr effect \cite{nandkishore11prl}, the helical ordered state exhibits nontrivial Josephson coupling with trivial superconductors \cite{chung13prb}. Whether the characteristics of these correlated phases can be probed in the experimental twisted bilayer graphene systems deserves further examination.

\begin{acknowledgments}
We thank Laura Classen and Liang Fu for fruitful discussions. We also acknowledge SangEun Han for pointing out an error in an equation. This research was sponsored by the Army Research Office and was accomplished under Grant No. W911NF-17-1-0482. The views and conclusions contained in this document are those of the authors and should not be interpreted as representing the official policies, either expressed or implied, of the Army Research Office or the U.S. Government. The U.S. Government is authorized to reproduce and distribute reprints for Government purposes notwithstanding any copyright notation herein.
\end{acknowledgments}


\appendix

\section{High-order Van Hove singularity}
\label{app:hovhs}

We discuss the general properties of high-order Van Hove singularity in this section. Our discussion focuses on a high-order saddle point $\mbf P$ with the general form of dispersion energy in its vicinity
\beeq
\ve_{\mbf P,\mbf k}\apx A_+k_+^{n_+}-A_-k_-^{n_-}.
\eneq
Here the prefactors $A_\pm>0$ are assumed, $\mbf k=(k_+,k_-)$ denotes the momentum deviation from $\mbf P$, and $n_\pm$ are positive even integers.

\subsection{Density of states}

The density of states $D(\ve)=\oint_{\txt{FS}_\ve,\mbf k}$ acquires a power-law divergence at the high-order saddle point $\mbf P$. Here $\txt{FS}_\ve$ represents the Fermi surface at the energy $\ve$. Assume an ultraviolet (UV) cutoff $\L$ for the dispersion energy and define the parameters $a_\pm=A_\pm k_\pm^{n_\pm}$
\beeq
D(\ve)
=\int_0^\L\prod_{s=\pm}\fr{2}{2\pi n_sA_s^{1/n_s}} da_sa_s^{1/n_s-1}\d(a_+-a_--\ve).
\eneq
Integrating out $a_+$ and setting $a=a_-$, the integral becomes
\beeq\beal
D(\ve)
&=\prod_{s=\pm}\fr{1}{\pi n_sA_s^{1/n_s}}\\&\hspace{11pt}\times\int_0^\L da(a+\ve)^{1/n_+-1}a^{1/n_--1}\t(a+\ve).
\enal\eneq
We separate the integrals for $\ve>0$ and $\ve<0$. With the reparametrization $u=a/|\ve|$ and the scaling dimension
\beeq
\e=1-\fr{1}{n_+}-\fr{1}{n_-},
\eneq
the integral reads
\beeq\beal
D(\ve)
&=\prod_{s=\pm}\fr{1}{\pi n_sA_s^{1/n_s}}|\ve|^{-\e}\Bigg[\t(\ve)\int_0^{\L/\ve} du\fr{u^{1/n_--1}}{(1+u)^{1/n_-+\e}}\\&\hspace{11pt}+\t(-\ve)\int_1^{\L/|\ve|} du\fr{(u-1)^{1/n_+-1}}{u^{1/n_++\e}}\Bigg].
\enal\eneq
Performing a further reparametrization $u'=u-1$ for the second integral and pushing the limit of UV cutoff to infinity $\L/|\ve|\rar\infty$, we rewrite the integral in terms of the Beta functions
\beeq\beal
D(\ve)
&=\prod_{s=\pm}\fr{1}{\pi n_sA_s^{1/n_s}}|\ve|^{-\e}\\&\hspace{11pt}\times[\t(\ve)B(1/n_-,\e)+\t(-\ve)B(1/n_+,\e)].
\enal\eneq
Using $B(x,y)=\G(x)\G(y)/\G(x+y)$ and $\G(1-z)\G(z)=\pi/\sin(\pi z)$, we arrive at the final form
\beeq
\label{eq:dospowerlawgeneral}
D(\ve)
=D_0\lf[\t(\ve)\sin\fr{\pi}{n_+}+\t(-\ve)\sin\fr{\pi}{n_-}\ri]|\ve|^{-\e}
\eneq
with the prefactor
\beeq
D_0=\fr{\G(\e)}{\pi}\prod_{s=\pm}\fr{\G(1/n_s)}{\pi n_sA_s^{1/n_s}}.
\eneq
Note that an asymmetry can generically be present on the two sides of the Van Hove doping.

\subsection{Susceptibility}

We now proceed to calculate the static susceptibilities (\ref{eq:pippph}) in the particle-hole (ph) and the particle-particle (pp) channels at the high-order Van Hove singularity. We focus particularly on the zero-momentum susceptibilities, while the finite-momentum ones depend generically on the structure of Fermi surface. The zero-momentum particle-hole susceptibility $\Pi^\txt{ph}_{\mbf 0}=\Pi^\txt{ph}_{\mbf q}|_{\mbf q\rar\mbf0}$ corresponds directly to the power-law divergent density of states (\ref{eq:dospowerlawgeneral}) $\Pi^\txt{ph}_{\mbf 0}=-\int d\ve D(\ve)\p_\ve n_F(\ve-\mu)$
\beeq\beal
\Pi^\txt{ph}_{\mbf 0}
&=\fr{D_0}{2}T^{-\e}\int dx|x|^{-\e}\cosh^{-2}\fr{x-\mu/T}{2}\\&\hspace{11pt}\times\fr{1}{2}\lf[\t(x)\sin\fr{\pi}{n_+}+\t(-x)\sin\fr{\pi}{n_-}\ri].
\enal\eneq
At $\mu=0$, we have
\beeq\beal
\Pi^\txt{ph}_{\mbf 0}
&=\fr{D_0}{2}T^{-\e}\int dx|x|^{-\e}\cosh^{-2}\fr{x}{2}\\&\hspace{11pt}\times\fr{1}{2}\lf[\t(x)\sin\fr{\pi}{n_+}+\t(-x)\sin\fr{\pi}{n_-}\ri].
\enal\eneq
Note that the power-law divergence is now controlled by the natural infrared (IR) cutoff $T$ with the same exponent $-\e$. Meanwhile, the zero-momentum particle-particle susceptibility enjoys the Cooper divergence
\beeq
\Pi^\txt{pp}_{\mbf 0}
=-\int d\ve D(\ve)\fr{n_F(\ve-\mu)-n_F(-[\ve-\mu])}{2(\ve-\mu)}.
\eneq
A direct evaluation leads to
\beeq\beal
\Pi^\txt{pp}_{\mbf 0}
&=\fr{D_0}{2}T^{-\e}\int dx|x|^{-\e}\fr{\tanh[(x-\mu/T)/2]}{x-\mu/T}\\&\hspace{11pt}\times\lf[\t(x)\sin\fr{\pi}{n_+}+\t(-x)\sin\fr{\pi}{n_-}\ri],
\enal\eneq
which becomes
\beeq\beal
\Pi^\txt{pp}_{\mbf 0}
&=\fr{D_0}{2}T^{-\e}\int dx|x|^{-\e}\fr{\tanh(x/2)}{x}\\&\hspace{11pt}\times\lf[\t(x)\sin\fr{\pi}{n_+}+\t(-x)\sin\fr{\pi}{n_-}\ri]
\enal\eneq
at $\mu=0$. Pushing the UV cutoff to infinity for the integral, an approximate relation with $\Pi^\txt{ph}_0$ \cite{isobe19prr} can be established by an integration by parts
\beeq
\Pi^\txt{pp}_{\mbf 0}
=\fr{1}{\e}\Pi^\txt{ph}_{\mbf 0}.
\eneq

\section{Renormalization group equations}
\label{app:rg}

\begin{figure}[t]
\centering
\includegraphics[scale = 1]{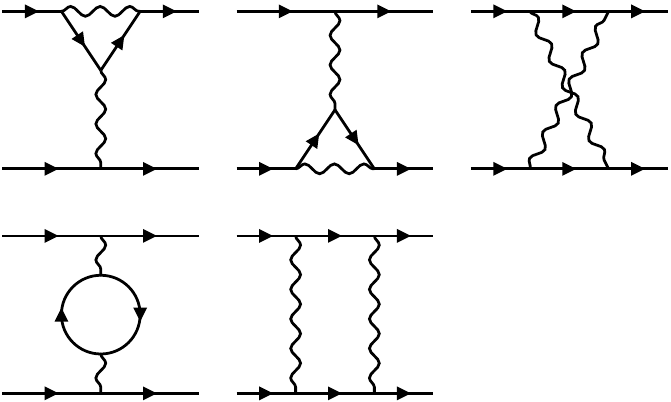}
\caption{\label{fig:1lc} One-loop corrections to the interactions.}
\end{figure}

\begin{figure*}[t]
\centering
\includegraphics[scale = 1]{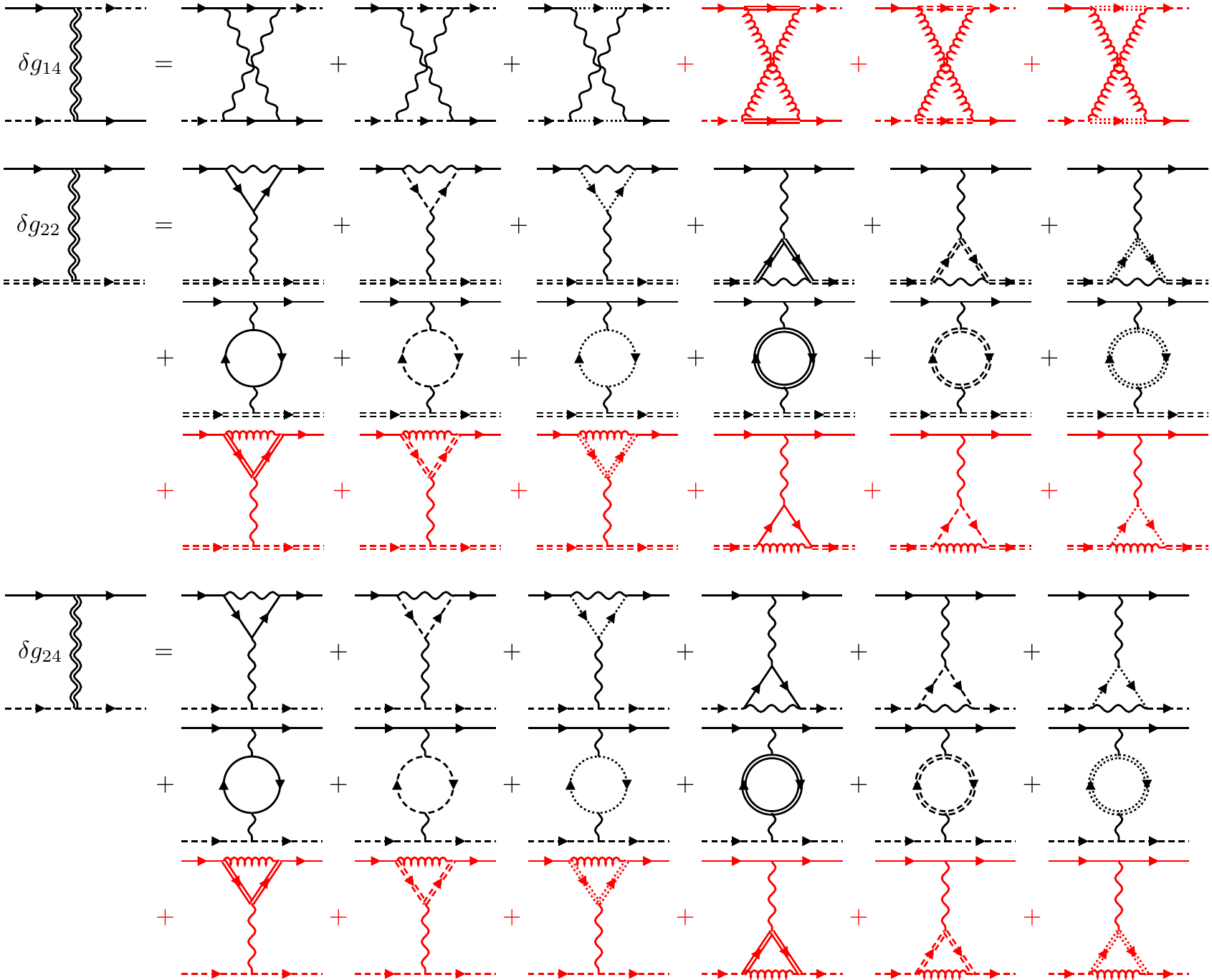}
\caption{\label{fig:rgeqs1} RG equations of the primary interactions without intervalley exchange (Part one). The red diagrams are those involve the intervalley exchange.}
\end{figure*}

\begin{figure*}[t]
\centering
\includegraphics[scale = 1]{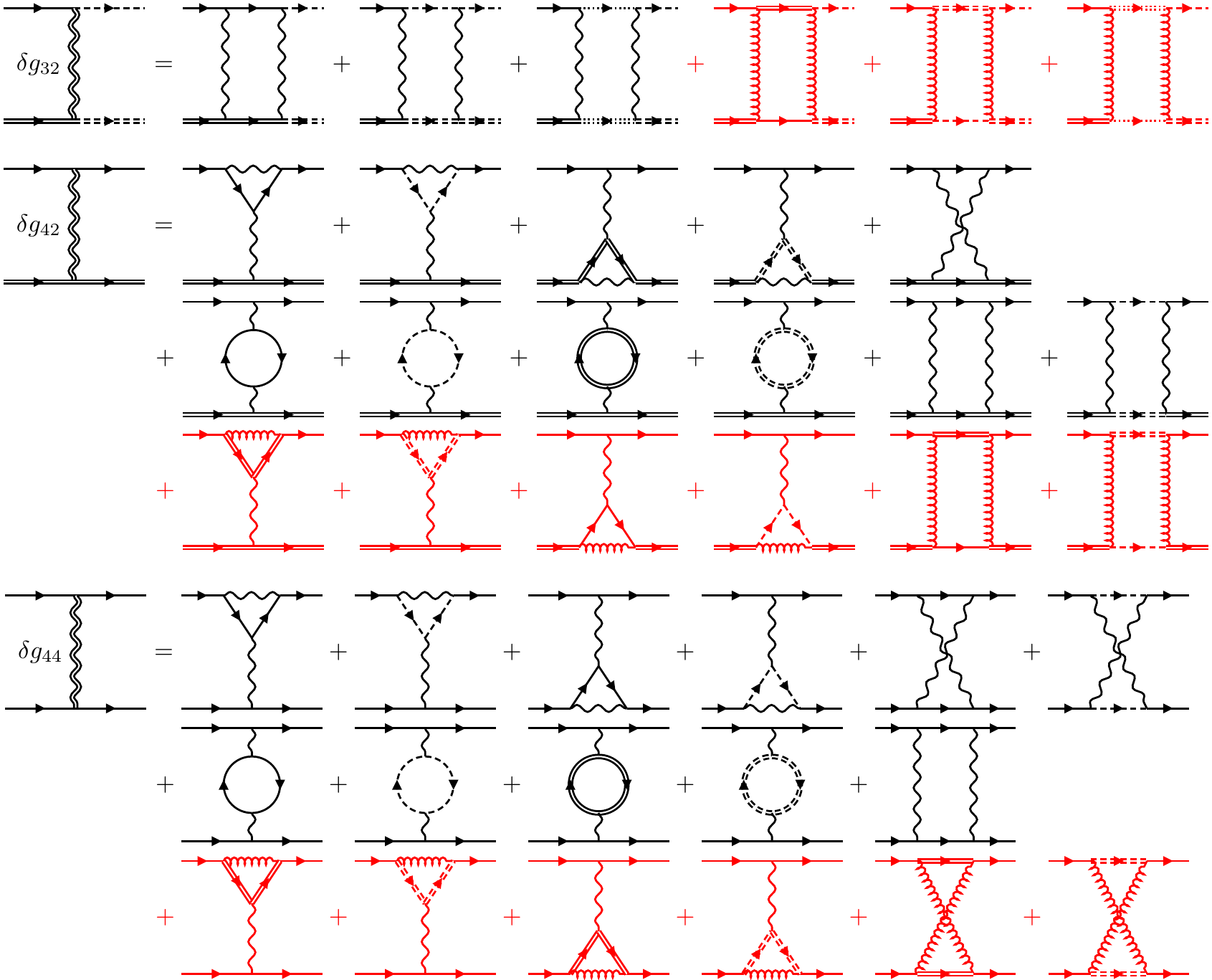}
\caption{\label{fig:rgeqs2} RG equations of the primary interactions without intervalley exchange (Part two). The red diagrams are those involve the intervalley exchange.}
\end{figure*}

\begin{figure*}[t]
\centering
\includegraphics[scale = 1]{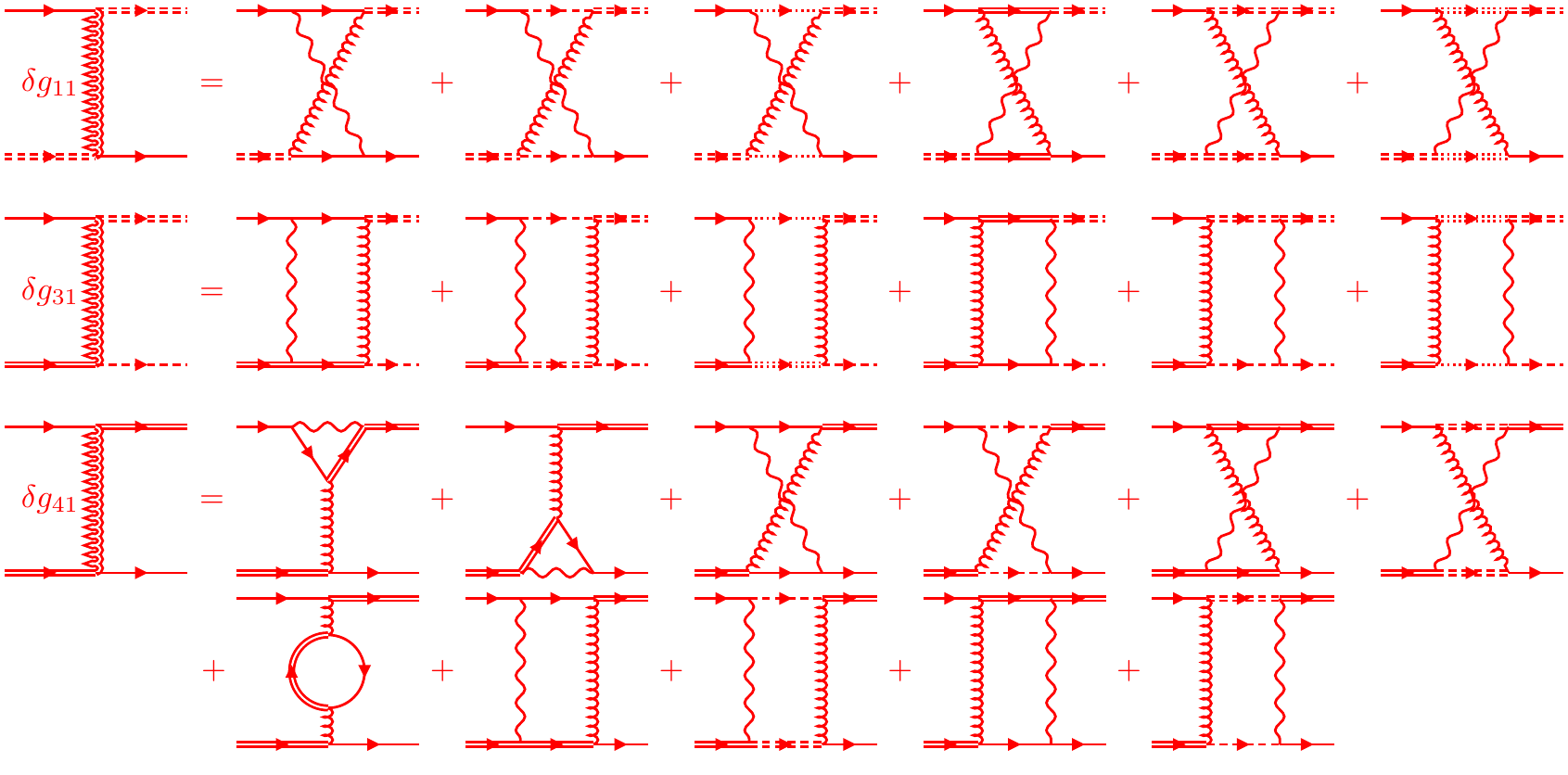}
\caption{\label{fig:rgeqswive} RG equations of the interactions with intervalley exchange. The red diagrams are those involve the intervalley exchange.}
\end{figure*}

In this section, we derive the renormalization group (RG) equations for the nine eligible interactions in the patch model. Consider the one-loop corrections to the interactions with decreasing temperature $T\rar0$ (Fig.~\ref{fig:1lc}). The according RG equations for the primary interactions without intervalley exchange read (Fig.~\ref{fig:rgeqs1})
\beeq
\label{eq:rgapp1}
\beal
\dot\l_{14}&=\e\l_{14}+d^\txt{ph}_{\mbf0}\l_{14}(\l_{14}+2\l_{44})\\&\hspace{11pt}+\big\{d^\txt{ph}_{\mbf0}\l_{11}(\l_{11}+2\l_{41})\big\}_\txt{wIVE},\\
\dot\l_{22}
&=\e\l_{22}+d^\txt{ph}_{\mbf0}[2\l_{22}(\l_{14}-2\l_{24}-\l_{44})\\&\hspace{11pt}+2\l_{42}(\l_{14}-2\l_{24})]\\&\hspace{11pt}+\big\{2d^\txt{ph}_{\mbf0}(\l_{11}\l_{24}+\l_{11}\l_{44}+\l_{24}\l_{41})\big\}_\txt{wIVE},\\
\dot\l_{24}
&=\e\l_{24}+d^\txt{ph}_{\mbf0}[2\l_{22}(-\l_{22}-2\l_{42})\\&\hspace{11pt}+2\l_{24}(\l_{14}-\l_{24}-\l_{44})+2\l_{14}\l_{44}]\\&\hspace{11pt}+\big\{2d^\txt{ph}_{\mbf0}(\l_{11}\l_{22}+\l_{11}\l_{42}+\l_{22}\l_{41})\big\}_\txt{wIVE},
\enal\eneq
and (Fig.~\ref{fig:rgeqs2})
\beeq
\label{eq:rgapp2}
\beal
\dot\l_{32}&=\e\l_{32}-\l_{32}(\l_{32}+2\l_{42})\\&\hspace{11pt}+\big\{-\l_{31}(\l_{31}+2\l_{41})\big\}_\txt{wIVE},\\
\dot\l_{42}
&=\e\l_{42}+d^\txt{ph}_{\mbf0}[4\l_{22}(\l_{14}-2\l_{24})-2\l_{42}\l_{44}]\\&\hspace{11pt}+d^\txt{ph}_{\mbf Q^o}\l_{42}^2-(2\l_{32}^2+\l_{42}^2)\\&\hspace{11pt}+\big\{2d^\txt{ph}_{\mbf0}(2\l_{11}\l_{24}+\l_{41}\l_{44})-(2\l_{31}^2+\l_{41}^2)\big\}_\txt{wIVE},\\
\dot\l_{44}
&=\e\l_{44}+d^\txt{ph}_{\mbf0}[2\l_{14}(\l_{14}+2\l_{24})-4\l_{22}^2-4\l_{24}^2\\&\hspace{11pt}-2\l_{42}^2+\l_{44}^2]-d^\txt{pp}_{-\mbf Q^o}\l_{44}^2\\&\hspace{11pt}+\big\{d^\txt{ph}_{\mbf0}[2\l_{11}(\l_{11}+2\l_{22})+\l_{41}(\l_{41}+2\l_{42})]\big\}_\txt{wIVE}.
\enal\eneq
The curly brackets indicate the corrections with intervalley exchange (wIVE), and are not included in the analysis with only primary interactions (Sec.~\ref{sec:rg}). Meanwhile, the corrections to the interactions with intervalley exchange take the form (Fig.~\ref{fig:rgeqswive})
\beeq
\label{eq:rgappwive}
\beal
\dot \l_{11}&=\e\l_{11}+2d^\txt{ph}_{\mbf0}(\l_{11}\l_{14}+\l_{11}\l_{44}+\l_{14}\l_{41}),\\
\dot \l_{31}&=\e\l_{31}-2(\l_{31}\l_{32}+\l_{31}\l_{42}+\l_{32}\l_{41}),\\
\dot \l_{41}
&=\e\l_{41}+2d^\txt{ph}_{\mbf0}(2\l_{11}\l_{14}+\l_{41}\l_{44})\\&\hspace{11pt}+2d^\txt{ph}_{\mbf Q^o}\l_{41}(-\l_{41}+\l_{42})-2(2\l_{31}\l_{32}+\l_{41}\l_{42}).
\enal\eneq
The whole set of RG equations (\ref{eq:rgapp1}), (\ref{eq:rgapp2}), and (\ref{eq:rgappwive}) is included in the analysis with intervalley exchange (Sec.~\ref{sec:wive}).

\section{Test vertex analysis}
\label{app:testvertex}

\begin{figure}[t]
\centering
\includegraphics[scale = 1]{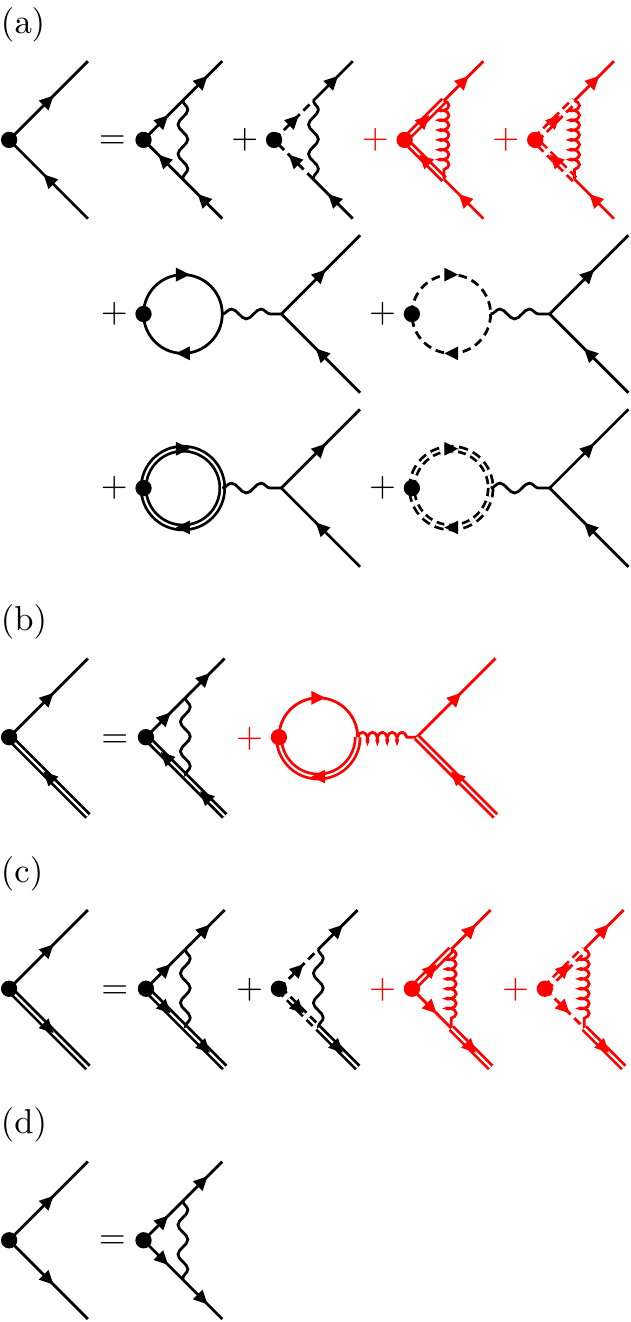}
\caption{\label{fig:tvflow} The corrections to the test vertices under RG in the particle-hole [(a) and (b)] and particle-particle [(c) and (d)] channels with zero-momentum [(a) and (c)] and finite-momentum $\mbf Q^o$ [(b) and (d)] pairings. The red diagrams are those involve the intervalley exchange. For the particle-hole channels [(a) and (b)], the diagrams with internal fermion loops are involved only for equal-spin pairings.}
\end{figure}

We conduct the test vertex analysis in the irreducible pairing channels which can receive the leading power-law divergence. These include the pairing channels at momenta $\mbf0$ and $\mbf Q^o$ in both particle-hole and particle-particle branches. Introducing the perturbing Hamiltonian (\ref{eq:dh}) with infinitesimal test vertices, we identify the one-loop corrections to the test vertices under RG (Fig.~\ref{fig:tvflow}). Such procedure is captured by a set of differential equations. The diagonalization of such equations indicates the irreducible pairing channels and the interactions therein.

\subsection{Particle-hole channels}

\subsubsection{Zero-momentum equal-spin pairings}

We first consider the test vertices involving zero-momentum particle-hole pairings with equal spin. The perturbing Hamiltonian (\ref{eq:dh}) reads
\beeq
\d H=\sum_{\a\tau\s}\D_{\a\tau\s}\psi_{\a\tau\s}^\dag\psi_{\a\tau\s}
\eneq
with real test vertices $\D_{\a\s\tau}$. The test vertices receive the corrections from the zero-momentum particle-hole susceptibility $\Pi^\txt{ph}_{\mbf0}$ under RG [Fig.~\ref{fig:tvflow}(a)]. This procedure is captured by the equation
\beeq\beal
\dot\D_{\a\tau\s}&=-d^\txt{ph}_{\mbf0}\Bigg[-\l_{44}\D_{\a\tau\s}-\l_{14}\sum_{\b\neq\a}\D_{\b\tau\s}\\&\hspace{11pt}+\Bigg\{\sum_{\tau\neq\tau'}\Bigg(-\l_{41}\D_{\a\tau'\s}-\l_{11}\sum_{\b\neq\a}\D_{\b\tau'\s}\Bigg)\Bigg\}_\txt{wIVE}\\&\hspace{11pt}+\sum_{\s'}\Bigg(\l_{44}\D_{\a\tau\s'}+\l_{24}\sum_{\b\neq\a}\D_{\b\tau\s'}\\&\hspace{11pt}+\sum_{\tau\neq\tau'}\Bigg[\l_{42}\D_{\a\tau'\s'}+\l_{22}\sum_{\b\neq\a}\D_{\b\tau'\s'}\Bigg]\Bigg)\Bigg].
\enal\eneq
The diagonalization in the patch sector identifies two pairing channels  $l=0,1$ with different patch orders $d_0$ and $d_{1,2}$, respectively
\beeq\beal
\dot\D_{l\tau\s}&=-d^\txt{ph}_{\mbf0}\Bigg[\l_l^1\D_{l\tau\s}+\Bigg\{\l_l^4\sum_{\tau\neq\tau'}\D_{l\tau'\s}\Bigg\}_\txt{wIVE}\\&\hspace{11pt}+\sum_{\s'}\Bigg(\l_l^2\D_{l\tau\s'}+\l_l^3\sum_{\tau\neq\tau'}\D_{l\tau'\s'}\Bigg)\Bigg].
\enal\eneq
Here the interactions are defined as
\beeq\beal
\l_0^1&=-2\l_{14}-\l_{44},\quad
\l_1^1=\l_{14}-\l_{44},\\
\l_0^2&=2\l_{24}+\l_{44},\quad
\l_1^2=-\l_{24}+\l_{44},\\
\l_0^3&=2\l_{22}+\l_{42},\quad
\l_1^3=-\l_{22}+\l_{42},\\
\big\{\l_0^4&=-2\l_{11}-\l_{41},\quad
\l_1^4=\l_{11}-\l_{41}\big\}_\txt{wIVE}.
\enal\eneq
We next diagonalize the equation in the spin sector, leading to two pairing channels $s=0,1$ with spin singlet and triplet pairings, respectively
\beeq
\dot\D_{l\tau s}=-d^\txt{ph}_{\mbf0}\Bigg(\l_{ls}^1\D_{l\tau s}+\l_{ls}^2\sum_{\tau\neq\tau'}\D_{l\tau's}\Bigg).
\eneq
The interactions in the equation now read
\beeq\beal
\l_{l0}^1&=2\l_l^2+\l_l^1,\quad
\l_{l1}^1=\l_l^1,\\
\l_{l0}^2&=2\l_l^3+\big\{\l_l^4\big\}_\txt{wIVE},\quad
\l_{l1}^2=\big\{\l_l^4\big\}_\txt{wIVE}.
\enal\eneq
Finally, we diagonalize the equation in the valley sector and get two pairing channels $v=e,o$ with even and odd valley pairings, respectively. The equation takes the form (\ref{eq:tvflowipc})
\beeq
\dot\D_{lvs}=-d^\txt{ph}_{\mbf0}\l_{lvs}\D_{lvs}
\eneq
with the interactions
\beeq
\l_{le\txt{/}os}=\pm\l_{ls}^2+\l_{ls}^1.
\eneq

There are eight irreducible pairing channels in total. The first ones are $s$- and $d$-wave Pomeranchuk orders ($s$/$d$POM), as well as $f$- and $p$-wave valley-polarized orders ($f$/$p$VP). The interactions $\l_{s\txt{POM/}f\txt{VP}}=\l_{0e/o0}$ and $\l_{d\txt{POM/}p\txt{VP}}=\l_{1e/o0}$ are derived as
\beeq\beal
\l_{s\txt{POM/}f\txt{VP}}
&=-2\l_{14}\pm4\l_{22}+4\l_{24}\pm2\l_{42}+\l_{44}\\&\hspace{11pt}+\big\{\mp2\l_{11}\mp\l_{41}\big\}_\txt{wIVE},\\
\l_{d\txt{POM/}p\txt{VP}}
&=\l_{14}\mp2\l_{22}-2\l_{24}\pm2\l_{42}+\l_{44}\\&\hspace{11pt}+\big\{\pm\l_{11}\mp\l_{41}\big\}_\txt{wIVE}.
\enal\eneq
Meanwhile, there are $s$- and $d$-wave ferromagnetisms ($s$/$d$FM), as well as $f$- and $p$-wave spin-valley-polarized orders ($f$/$p$SVP). The interactions $\l_{s\txt{FM/}f\txt{SVP}}=\l_{0e/o1}$ and $\l_{d\txt{FM/}p\txt{SVP}}=\l_{1e/o1}$ take the form
\beeq\label{eq:intfmsvp}\beal
\l_{s\txt{FM/}f\txt{SVP}}&=-2\l_{14}-\l_{44}+\big\{\mp2\l_{11}\mp\l_{41}\big\}_\txt{wIVE},\\
\l_{d\txt{FM/}p\txt{SVP}}&=\l_{14}-\l_{44}+\big\{\pm\l_{11}\mp\l_{41}\big\}_\txt{wIVE}.
\enal\eneq

\subsubsection{Zero-momentum opposite-spin pairings}

We next consider the zero-momentum particle-hole pairings with opposite spins. The perturbing Hamiltonian reads
\beeq
\d H=\sum_{\a\tau,\s>\s'}(\D_{\a\tau\s\s'}\psi_{\a\tau\s}^\dag\psi_{\a\tau\s'}+\txt{H.c.}),
\eneq
where the test vertices $\D_{\a\tau\s\s'}$ receive corrections from the zero-momentum particle-hole susceptibilities $\Pi^\txt{ph}_{\mbf0}$ under RG [Fig.~\ref{fig:tvflow}(a)]. Note that the diagrams with internal fermion loops are not involved in the corrections. The according equation reads
\beeq\beal
&\dot\D_{\a\tau\s\s'}=-d^\txt{ph}_{\mbf0}\Bigg[-\l_{44}\D_{\a\tau\s\s'}-\l_{14}\sum_{\b\neq\a}\D_{\b\tau\s\s'}\\&\hspace{11pt}+\Bigg\{\sum_{\tau\neq\tau'}\Bigg(-\l_{41}\D_{\a\tau'\s\s'}-\l_{11}\sum_{\b\neq\a}\D_{\b\tau'\s\s'}\Bigg)\Bigg\}_\txt{wIVE}\Bigg].
\enal\eneq
The diagonalization in the patch sector identifies two pairing channels $l=0,1$ with different patch orders $d_0$ and $d_{1,2}$, respectively
\beeq\beal
\dot\D_{l\tau\s\s'}&=-d^\txt{ph}_{\mbf0}\Bigg(\l_l^1\D_{l\tau\s\s'}+\Bigg\{\l_l^2\sum_{\tau\neq\tau'}\D_{l\tau'\s\s'}\Bigg\}_\txt{wIVE}\Bigg).
\enal\eneq
Here the interactions read
\beeq\beal
\l_0^1&=-2\l_{14}-\l_{44},\quad
\l_1^1=\l_{14}-\l_{44},\\
\big\{\l_0^2&=-2\l_{11}-\l_{41},\quad
\l_1^2=\l_{11}-\l_{41}\big\}_\txt{wIVE}.
\enal\eneq
A further diagonalization in the valley sector finds two pairing channels $v=e,o$ with even and odd valley pairings, respectively
\beeq
\dot\D_{lv\s\s'}=-d^\txt{ph}_{\mbf0}\l_{lv}\D_{lv\s\s'}.
\eneq
The interactions in these two channels are
\beeq
\l_{le\txt{/}o}=\l_l^1+\big\{\pm\l_l^2\big\}_\txt{wIVE}.
\eneq

The four irreducible pairing channels correspond to the $s$- and $d$-wave ferromagnetisms, as well as the $f$- and $p$-wave spin-valley-polarized orders. Note that the interactions $\l_{s\txt{FM/}f\txt{SVP}}=\l_{0e/o}$ and $\l_{d\txt{FM/}p\txt{SVP}}=\l_{1e/o}$ are consistent with the results (\ref{eq:intfmsvp}) from the equal-spin pairings.

\subsubsection{Charge and spin density waves}

With the test vertices coupled to the particle-hole pairings at finite momenta $\mbf Q^o$'s, the perturbing Hamiltonian
\beeq
\d H=\sum_{\a,\tau<\tau',\s\s'}(\D_{\a\tau\s\s'}\psi_{\a\tau'\s'}^\dag\psi_{\a\tau\s}+\txt{H.c.})
\eneq
is introduced. The test vertices receive the corrections from the finite-momentum particle-hole susceptibilities $\Pi^\txt{ph}_{\mbf Q^o}$ under RG [Fig.~\ref{fig:tvflow}(b)]. For the equal-spin pairings with $\s=\s'$, the corrections are described by the equation
\beeq\beal
\dot\D_{\a\tau\s}&=-d^\txt{ph}_{\mbf Q^o}\Bigg(-\l_{42}\D_{\a\tau\s}+\Bigg\{\sum_{\s'}\l_{41}\D_{\a\tau\s'}\Bigg\}_\txt{wIVE}\Bigg).
\enal\eneq
A diagonalization in the spin sector finds the spin singlet and triplet solutions $s=0,1$
\beeq
\dot\D_{\a\tau s}=-d^\txt{ph}_{\mbf Q^o}\l_s\D_{\a\tau s}.
\eneq
These two solutions correspond to the charge and spin density wave channels (C/SDW$^o$), respectively. With $\l_\txt{CDW$^o$}=\l_0$ and $\l_\txt{SDW$^o$}=\l_1$, we find
\beeq\label{eq:intsdw}\beal
\l_\txt{CDW$^o$}&=-\l_{42}+\big\{2\l_{41}\big\}_\txt{wIVE},\\
\l_\txt{SDW$^o$}&=-\l_{42}.
\enal\eneq

For the opposite-spin pairings with $\s\neq\s'$, the diagrams with internal fermion loops are not involved in the corrections
\beeq\beal
\dot\D_{\a\tau\s\s'}&=-d^\txt{ph}_{\mbf Q^o}(-\l_{42})\D_{\a\tau\s\s'}.
\enal\eneq
The solution to this equation corresponds to the spin density wave channel, and the interaction is consistent with the result (\ref{eq:intsdw}) from the equal-spin pairings.

\subsection{Particle-particle channels}

\subsubsection{Superconducting channels}

For the zero-momentum particle-particle pairing channels, the perturbing Hamiltonian (\ref{eq:dh}) induced by the test vertices is
\beeq
\d H=\sum_{\a,\tau\neq\tau',\s\s'}(\D_{\a\tau\s\s'}\psi_{\a\tau\s}^\dag\psi_{\a\tau'\s'}^\dag+\txt{H.c.}).
\eneq
The test vertices receive the corrections from the zero-momentum particle-particle susceptibility $\Pi^\txt{pp}_{\mbf0}$ under RG [Fig.~\ref{fig:tvflow}(c)], as captured by the equation
\beeq\beal
&\dot\D_{\a\tau\s\s'}=-\Bigg[\l_{42}\D_{\a\tau\s\s'}+\l_{32}\sum_{\b\neq\a}\D_{\b\tau\s\s'}\\&\hspace{11pt}+\Bigg\{\sum_{\tau'\neq\tau}\Bigg(\l_{41}\D_{\a\tau'\s\s'}+\l_{31}\sum_{\b\neq\a}\D_{\b\tau'\s\s'}\Bigg)\Bigg\}_\txt{wIVE}\Bigg].
\enal\eneq
The diagonalization in the patch sector identifies two pairing channels $l=0,1$ with different patch orders $d_0$ and $d_{1,2}$, respectively
\beeq
\dot\D_{l\tau\s\s'}=-\Bigg(\l_l^1\D_{l\tau\s\s'}+\Bigg\{\l_l^2\sum_{\tau'\neq\tau}\D_{\a\tau'\s\s'}\Bigg\}_\txt{wIVE}\Bigg).
\eneq
The interactions here are derived as
\beeq\beal
\l_0^1&=2\l_{32}+\l_{42},\quad
\l_1^1=-\l_{32}+\l_{42},\\
\big\{\l_0^2&=2\l_{31}+\l_{41},\quad
\l_1^2=-\l_{31}+\l_{41}\big\}_\txt{wIVE}.
\enal\eneq
A further diagonalization in the valley sector uncovers two pairing channels $v=e,o$ with even and odd valley pairings, respectively
\beeq
\dot\D_{lv\s\s'}=-\l_{lv}\D_{lv\s\s'},
\eneq
and the interactions are
\beeq
\l_{le\txt{/}o}=\l_l^1+\big\{\pm\l_l^2\big\}_\txt{wIVE}.
\eneq
We thus identify the four superconducting channels with $s$-, $f$-, $d$-, and $p$-wave orders ($s$/$f$/$d$/$p$SC). The interactions $\l_{s\txt{/}f\txt{SC}}=\l_{0e/o}$ and $\l_{d\txt{/}p\txt{SC}}=\l_{1e/o}$ are derived as
\beeq\beal
\l_{s\txt{/}f\txt{SC}}&=2\l_{32}+\l_{42}+\big\{\pm2\l_{31}\pm\l_{41}\big\}_\txt{wIVE},\\
\l_{d\txt{/}p\txt{SC}}&=-\l_{32}+\l_{42}+\big\{\mp\l_{31}\pm\l_{41}\big\}_\txt{wIVE}.
\enal\eneq

\begin{figure*}[t]
\centering
\includegraphics[scale = 1]{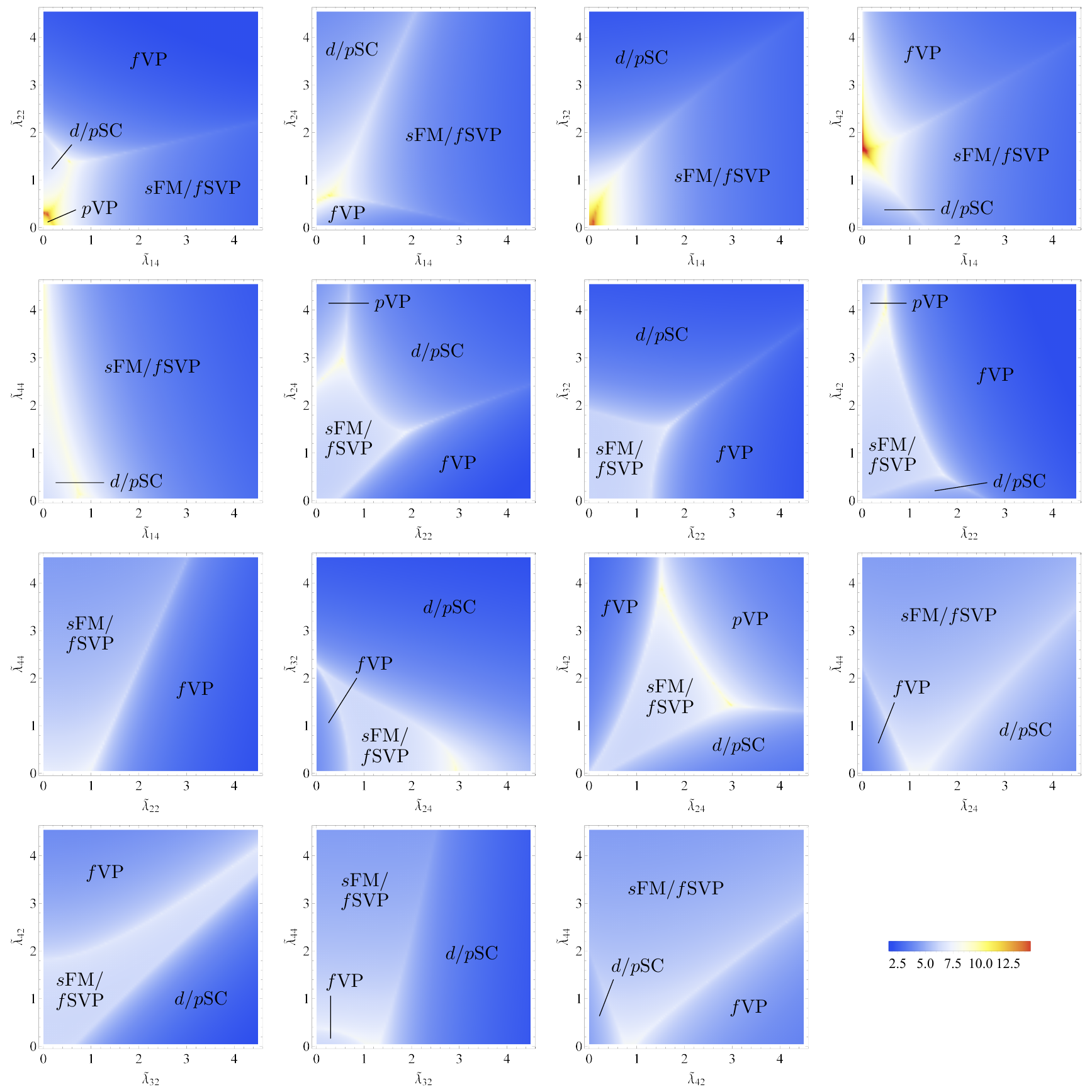}
\caption{\label{fig:pdall} The full set of two-interaction phase diagrams.}
\end{figure*}

\subsubsection{Pair density waves}

With the test vertices coupled to the particle-particle pairings at finite momenta $\mbf Q^o$'s, the perturbing Hamiltonian
\beeq
\d H=\sum_{\a\tau\s\s'}(\D_{\a\tau\s\s'}\psi_{\a\tau\s'}^\dag\psi_{\a\tau\s}^\dag+\txt{H.c.})
\eneq
is introduced. The test vertices receive the corrections from the finite-momentum particle-particle susceptibilities $\Pi^\txt{pp}_{\mbf Q^o}$ under RG [Fig.~\ref{fig:tvflow}(d)], which is described by the equation
\beeq
\dot\D_{\a\tau\s\s'}=-d^\txt{pp}_{-\mbf Q^o}\l_{44}\D_{\a\tau\s\s'}.
\eneq
The interaction in the according pair density wave channels (PDW$^o$) is thus identified as
\beeq
\l_\txt{PDW$^o$}=\l_{44}.
\eneq

\section{Phase diagram}
\label{app:pd}

In Sec.~\ref{sec:inst}, we have shown selected two-interaction phase diagrams of the potential instabilities from the weakly repulsive primary interactions (Fig.~\ref{fig:pd}). Here we list the full set of two-interaction phase diagrams (Fig.~\ref{fig:pdall}), from which the features discussed in Sec.~\ref{sec:inst} may be more easily observed.

\section{Ginzburg-Landau free energy}
\label{app:gl}

Our RG analysis has uncovered three potential instabilities with degenerate structures. These include the $p$-wave superconductivity, the $d$-wave superconductivity, and the $p$-wave valley-polarized order. The degeneracy breakdown in these degenerate channels can be solved by the Ginzburg-Landau analysis. By expanding the free energy with respect to the infinitesimal order parameters near the critical temperature, the energetically favored ground states are determined through the free energy minimization. Here we derive the expanded free energy near the critical temperature, known as the Ginzburg-Landau free energy. The results are then adopted in the analysis in Sec.~\ref{sec:order}, where the energetically favored ground states are identified.

\subsection{$p$-wave superconductivity}

We project the interacting model onto the $p$-wave superconducting channel. The action reads
\beeq
S=\int_\tau\lf[\sum_\k\psi_\k^\dag(\p_\tau+\xi_\k)\psi_\k+\fr{g_{p\txt{SC}}}{2}(\mbf P_1^\dag\cdot\mbf P_1+\mbf P_2^\dag\cdot\mbf P_2)\ri],
\eneq
where $\tau$ is the imaginary time and the pairing operator
\beeq
\mbf P_a^\dag=\psi_+^\dag\lf(\fr{\bsb\s}{\sqrt2}\ri)d_a[i(i\s^2)(\psi_-^\dag)^T]
\eneq
is defined. We have reduced the irreducible valley pairing representations and take $\psi_\pm$ as in the $\k=\pm$ valleys, respectively. The interaction is taken negative $g_{p\txt{SC}}<0$. We conduct a Hubbard-Stratonovich transformation, where the pairing operators are decoupled by the bosonic complex vector order parameters $\mbf\D_{1,2}$. Impose the static condition for the order parameters $\mbf\D_a(\tau)=\mbf\D_a$. Defining the Nambu spinor $\Psi=(\psi_+,i[i\s^2][\psi_-^\dag]^T)^T$ and integrating it out in the Matsubara frequency representation $\Psi(\tau)=\sqrt T\sum_n\Psi_ne^{-i\o_n\tau}$, we arrive at the mean-field free energy
\beeq
f=\fr{2}{|g_{p\txt{SC}}|}(|\mbf\D_1|^2+|\mbf\D_2|^2)-\Tr\ln(-\mca G^{-1}).
\eneq
The inverse Gor'kov Green's function has been defined
\beeq
\mca G^{-1}=\lf(\bear{cc}G_+^{-1}&\sum_a\mbf\D_a\cdot[\bsb\s/\sqrt2]d_a\\\sum_a\mbf{\bar\D}_a\cdot[\bsb\s/\sqrt2]d_a&G_-^{-1}\enar\ri),
\eneq
where the free electron and hole propagators are $G_\pm=[i\o_n\mp(\ve_\pm-\mu)]^{-1}$.

We expand the free energy with respect to the infinitesimal order parameters near the critical temperature $T_c$. Define $\mca G_0=\mca G_0(\mbf\D_{1,2}=0)$ and $\hat\D=\mca G^{-1}-\mca G_0^{-1}$. Ignoring the constant part of the free energy, we perform the expansion up to quartic order
\beeq
f=\fr{2}{|g_{p\txt{SC}}|}(|\mbf\D_1|^2+|\mbf\D_2|^2)+\fr{1}{2}\Tr(\mca G_0\hat\D)^2+\fr{1}{4}\Tr(\mca G_0\hat\D)^4.
\eneq
Here the infinitesimal expansion parameter reads
\beeq\beal
(\mca G_0\hat\D)^2
&=\fr{1}{2}G_+G_-\sum_{ab}d_ad_b\\&\hspace{11pt}\times\txt{diag}([\mbf\D_a\cdot\bsb\s][\mbf{\bar\D}_b\cdot\bsb\s],[\mbf{\bar\D}_a\cdot\bsb\s][\mbf\D_b\cdot\bsb\s]).
\enal\eneq
The quadratic-order terms in the free energy are
\beeq\beal
f^{(2)}
&=\fr{2}{|g_{p\txt{SC}}|}(|\mbf\D_1|^2+|\mbf\D_2|^2)+\fr{1}{4}\Tr\Bigg[G_+G_-\sum_{ab}d_ad_b\\&\hspace{11pt}\times\txt{diag}([\mbf\D_a\cdot\bsb\s][\mbf{\bar\D}_b\cdot\bsb\s],[\mbf{\bar\D}_a\cdot\bsb\s][\mbf\D_b\cdot\bsb\s])\Bigg].
\enal\eneq
Utilizing $\Tr(\s^i\s^j)=2\d_{ij}$ and $\Tr(d_ad_b)=\d_{ab}$, we obtain
\beeq
f^{(2)}
=\lf[\fr{2}{|g_{p\txt{SC}}|}+\Tr(G_+G_-)\ri](|\mbf\D_1|^2+|\mbf\D_2|^2).
\eneq
The square bracket term takes the form $T-T_c$ and turns negative below $T_c$. Meanwhile, the quartic order terms read
\beeq\beal
f^{(4)}
&=\fr{1}{16}\Tr\Bigg[G_+^2G_-^2\sum_{abcd}d_ad_bd_cd_d\\&\hspace{11pt}\times\txt{diag}([\mbf\D_a\cdot\bsb\s][\mbf{\bar\D}_b\cdot\bsb\s][\mbf\D_c\cdot\bsb\s][\mbf{\bar\D}_d\cdot\bsb\s],\\&\hspace{11pt}[\mbf{\bar\D}_a\cdot\bsb\s][\mbf\D_b\cdot\bsb\s][\mbf{\bar\D}_c\cdot\bsb\s][\mbf\D_d\cdot\bsb\s])\Bigg],
\enal\eneq
where the term $\Tr(G_+^2G_-^2)$ can be verified to be positive. The nonvanishing traces in the patch sector are $\Tr(d_1^4)=\Tr(d_2^4)=1/2$ and $\Tr(d_1^2d_2^2)=\Tr(d_1d_2d_1d_2)=1/6$. This implies
\beeq\beal
f^{(4)}
&=\fr{1}{16}\Tr(G_+^2G_-^2)\\&\hspace{11pt}\times\Bigg\{\sum_a\Tr(\mbf\D_a\cdot\bsb\s)(\mbf{\bar\D}_a\cdot\bsb\s)(\mbf\D_a\cdot\bsb\s)(\mbf{\bar\D}_a\cdot\bsb\s)\\&\quad\,+\fr{1}{3}\sum_{a\neq b}[
\Tr(\mbf\D_a\cdot\bsb\s)(\mbf{\bar\D}_a\cdot\bsb\s)(\mbf\D_b\cdot\bsb\s)(\mbf{\bar\D}_b\cdot\bsb\s)\\&\hspace{11pt}+\Tr(\mbf\D_a\cdot\bsb\s)(\mbf{\bar\D}_b\cdot\bsb\s)(\mbf\D_b\cdot\bsb\s)(\mbf{\bar\D}_a\cdot\bsb\s)\\&\hspace{11pt}+\Tr(\mbf\D_a\cdot\bsb\s)(\mbf{\bar\D}_b\cdot\bsb\s)(\mbf\D_a\cdot\bsb\s)(\mbf{\bar\D}_b\cdot\bsb\s)]\Bigg\}.
\enal\eneq
In the spin sector, the nonvanishing terms are $\Tr(\s^i)^4=2$ and $\Tr[(\s^i)^2(\s^j)^2]=-\Tr(\s^i\s^j\s^i\s^j)=2$ for $i\neq j$, leading to
\beeq\beal
&\Tr[(\mbf\D_a\cdot\bsb\s)(\mbf{\bar\D}_b\cdot\bsb\s)(\mbf\D_c\cdot\bsb\s)(\mbf{\bar\D}_d\cdot\bsb\s)]\\
&=2\sum_i\D_{ai}\bar\D_{bi}\D_{ci}\bar\D_{di}+2\sum_{i\neq j}(\D_{ai}\bar\D_{bi}\D_{cj}\bar\D_{dj}\\&\hspace{11pt}-\D_{ai}\bar\D_{bj}\D_{ci}\bar\D_{dj}+\D_{ai}\bar\D_{bj}\D_{cj}\bar\D_{di}).
\enal\eneq
Adopting this result to the quartic terms with different patch configurations separately, we derive
\beeq\beal
&\Tr[(\mbf\D_a\cdot\bsb\s)(\mbf{\bar\D}_a\cdot\bsb\s)(\mbf\D_a\cdot\bsb\s)(\mbf{\bar\D}_a\cdot\bsb\s)]\\&\hspace{11pt}=2|\mbf\D_a|^4+2|\mbf{\bar\D}_a\times\mbf\D_a|^2,\\
&\Tr[(\mbf\D_a\cdot\bsb\s)(\mbf{\bar\D}_a\cdot\bsb\s)(\mbf\D_b\cdot\bsb\s)(\mbf{\bar\D}_b\cdot\bsb\s)]\\&\hspace{11pt}=2|\mbf\D_a|^2|\mbf\D_b|^2-2|\mbf\D_a\cdot\mbf\D_b|^2+2|\mbf\D_a\cdot\mbf{\bar\D}_b|^2,\\
&\Tr[(\mbf\D_a\cdot\bsb\s)(\mbf{\bar\D}_b\cdot\bsb\s)(\mbf\D_b\cdot\bsb\s)(\mbf{\bar\D}_a\cdot\bsb\s)]\\&\hspace{11pt}=2|\mbf\D_a|^2|\mbf\D_b|^2-2|\mbf\D_a\cdot\mbf\D_b|^2+2|\mbf\D_a\cdot\mbf{\bar\D}_b|^2,\\
&\Tr[(\mbf\D_a\cdot\bsb\s)(\mbf{\bar\D}_b\cdot\bsb\s)(\mbf\D_a\cdot\bsb\s)(\mbf{\bar\D}_b\cdot\bsb\s)]\\&\hspace{11pt}=4(\mbf\D_a\cdot\mbf{\bar\D}_b)^2-2\mbf\D_a^2\mbf{\bar\D}_b^2.
\enal\eneq
The quartic terms are then identified as
\beeq\beal
f^{(4)}
&=\fr{1}{8}\Tr(G_+^2G_-^2)\bigg\{(|\mbf\D_1|^2+|\mbf\D_2|^2)^2\\&\hspace{11pt}+|\mbf{\bar\D}_1\times\mbf\D_1|^2+|\mbf{\bar\D}_2\times\mbf\D_2|^2\\&\hspace{11pt}+\fr{1}{3}[-2|\mbf\D_1|^2|\mbf\D_2|^2-\mbf\D_1^2\mbf{\bar\D}_2^2-\mbf{\bar\D}_1^2\mbf\D_2^2\\&\hspace{11pt}+2(\mbf\D_1\cdot\mbf{\bar\D}_2)^2+2(\mbf{\bar\D}_1\cdot\mbf\D_2)^2\\&\hspace{11pt}-4|\mbf\D_1\cdot\mbf\D_2|^2+4|\mbf\D_1\cdot\mbf{\bar\D}_2|^2]\bigg\}.
\enal\eneq
With the identities
\beeq\beal
|\mbf\D_1\times\mbf\D_2|^2
&=|\mbf\D_1|^2|\mbf\D_2|^2-|\mbf\D_1\cdot\mbf{\bar\D}_2|^2,\\
|\mbf{\bar\D}_1\times\mbf\D_2|^2
&=|\mbf\D_1|^2|\mbf\D_2|^2-|\mbf\D_1\cdot\mbf\D_2|^2,\\
(\mbf\D_1\times\mbf{\bar\D}_2)^2
&=\mbf\D_1^2\mbf{\bar\D}_2^2-(\mbf\D_1\cdot\mbf{\bar\D}_2)^2,
\enal\eneq
the quartic terms can be reformulated as
\beeq\beal
f^{(4)}
&=\fr{1}{8}\Tr(G_+^2G_-^2)\bigg\{(|\mbf\D_1|^2+|\mbf\D_2|^2)^2\\&\hspace{11pt}+|\mbf{\bar\D}_1\times\mbf\D_1|^2+|\mbf{\bar\D}_2\times\mbf\D_2|^2\\&\hspace{11pt}+\fr{1}{3}[-2|\mbf\D_1|^2|\mbf\D_2|^2+\mbf\D_1^2\mbf{\bar\D}_2^2+\mbf{\bar\D}_1^2\mbf\D_2^2\\&\hspace{11pt}-2(\mbf\D_1\times\mbf{\bar\D}_2)^2-2(\mbf{\bar\D}_1\times\mbf\D_2)^2\\&\hspace{11pt}+4|\mbf{\bar\D}_1\times\mbf\D_2|^2-4|\mbf\D_1\times\mbf\D_2|^2]\bigg\}.
\enal\eneq
Combining these results, we obtain the Ginzburg-Landau free energy (\ref{eq:freeengpsc}) which is adopted in Sec.~\ref{sec:order}.

\subsection{$d$-wave superconductivity}

We now project the interacting model onto the $d$-wave superconducting channel. The action reads
\beeq
S=\int_\tau\lf[\sum_\k\psi_\k^\dag(\p_\tau+\xi_\k)\psi_\k+\fr{g_{d\txt{SC}}}{2}(P_1^\dag P_1+P_2^\dag P_2)\ri],
\eneq
where the pairing operator
\beeq
P_a^\dag=\psi_+^\dag\lf(\fr{\s^0}{\sqrt2}\ri)d_a[i(i\s^2)(\psi_-^\dag)^T]
\eneq
is defined. We have again reduced the irreducible valley pairing representations. The interaction is taken negative $g_{d\txt{SC}}<0$. We conduct a Hubbard-Stratonovich transformation, where the pairing operators are decoupled by the bosonic complex scalar order parameters $\D_{1,2}$. Impose the static condition for the order parameters $\D_a(\tau)=\D_a$. Defining the Nambu spinor $\Psi=(\psi_+,i[i\s^2][\psi_-^\dag]^T)^T$ and integrating it out in the Matsubara frequency representation $\Psi(\tau)=\sqrt T\sum_n\Psi_ne^{-i\o_n\tau}$, we arrive at the mean-field free energy
\beeq
f=\fr{2}{|g_{d\txt{SC}}|}(|\D_1|^2+|\D_2|^2)-\Tr\ln(-\mca G^{-1}).
\eneq
Here the inverse Gor'kov Green's function has been defined
\beeq
\mca G^{-1}=\lf(\bear{cc}G_+^{-1}&\sum_a\D_a[\s^0/\sqrt2]d_a\\\sum_a\bar\D_a[\s^0/\sqrt2]d_a&G_-^{-1}\enar\ri).
\eneq

We expand the free energy with respect to the infinitesimal order parameters near the critical temperature $T_c$. Define $\mca G_0=\mca G_0(\D_{1,2}=0)$ and $\hat\D=\mca G^{-1}-\mca G_0^{-1}$. Ignoring the constant part of the free energy, we perform the expansion up to quartic order
\beeq
f=\fr{2}{|g_{d\txt{SC}}|}(|\D_1|^2+|\D_2|^2)+\fr{1}{2}\Tr(\mca G_0\hat\D)^2+\fr{1}{4}\Tr(\mca G_0\hat\D)^4.
\eneq
Here the infinitesimal expansion parameter reads
\beeq\beal
(\mca G_0\hat\D)^2
&=\fr{1}{2}G_+G_-\sum_{ab}d_ad_b\\&\hspace{11pt}\times\txt{diag}([\D_a\s^0][\bar\D_b\s^0],[\bar\D_a\s^0][\D_b\s^0]).
\enal\eneq
The quadratic-order terms in the free energy are
\beeq\beal
f^{(2)}
&=\fr{2}{|g_{d\txt{SC}}|}(|\D_1|^2+|\D_2|^2)+\fr{1}{4}\Tr\Bigg[G_+G_-\sum_{ab}d_ad_b\\&\hspace{11pt}\times\txt{diag}([\D_a\s^0][\bar\D_b\s^0],[\bar\D_a\s^0][\D_b\s^0])\Bigg],
\enal\eneq
which are derived as
\beeq
f^{(2)}
=\lf[\fr{2}{|g_{d\txt{SC}}|}+\Tr(G_+G_-)\ri](|\D_1|^2+|\D_2|^2).
\eneq
The square bracket term takes the form $T-T_c$ and turns negative below $T_c$. Meanwhile, the quartic order terms read
\beeq\beal
f^{(4)}
&=\fr{1}{16}\Tr\Bigg[G_+^2G_-^2\sum_{abcd}d_ad_bd_cd_d\\&\hspace{11pt}\times\txt{diag}([\D_a\s^0][\bar\D_b\s^0][\D_c\s^0][\bar\D_d\s^0],\\&\hspace{11pt}[\bar\D_a\s^0][\D_b\s^0][\bar\D_c\s^0][\D_d\s^0])\Bigg],
\enal\eneq
which can be evaluated as
\beeq\beal
f^{(4)}
&=\fr{1}{8}\Tr(G_+^2G_-^2)\bigg[(|\D_1|^2+|\D_2|^2)^2\\&\quad\,+\fr{1}{3}
(-2|\D_1|^2|\D_2|^2+\D_1^2\bar\D_2^2+\bar\D_1^2\D_2^2)\bigg].
\enal\eneq
Combining these results, we obtain the Ginzburg-Landau free energy (\ref{eq:freeengdsc}) which is adopted in Sec.~\ref{sec:order}.

\subsection{$p$-wave valley-polarized order}

Projecting the interacting model onto the $p$-wave valley-polarized order channel, we have the action
\beeq
S=\int_\tau\lf[\psi^\dag(\p_\tau+\xi)\psi+\fr{g_{p\txt{VP}}}{2}\lf(P_1^2+P_2^2\ri)\ri].
\eneq
Here the interaction is taken negative $g_{p\txt{VP}}<0$ and the pairing operator
\beeq
P_a=\psi^\dag\lf(\fr{\tau^3}{\sqrt2}\ri)\lf(\fr{\s^0}{\sqrt2}\ri)d_a\psi
\eneq
is defined. We perform a Hubbard-Stratonovich transformation, where the pairing operators are decoupled by the bosonic real scalar order parameters $\D_{1,2}$. Impose the static condition for the order parameter $\D_a(\tau)=\D_a$. Integrating the fermions out in the Matsubara frequency representation $\psi(\tau)=\sqrt T\sum_n\psi_ne^{-i\o_n\tau}$, we arrive at the mean-field free energy
\beeq
f=\fr{2}{|g_{p\txt{VP}}|}(\D_1^2+\D_2^2)-\Tr\ln(-\mca G^{-1}).
\eneq
The inverse Green's function is defined
\beeq
\mca G^{-1}=G^{-1}+\sum_a\D_a\lf(\fr{\tau^3}{\sqrt2}\ri)\lf(\fr{\s^0}{\sqrt2}\ri)d_a
\eneq
with the free electron propagator $G=[i\o_n-(\ve-\mu)]^{-1}$.

We expand the free energy with respect to the infinitesimal order parameters near the critical temperature $T_c$. Define $\mca G_0=\mca G_0(\D_{1,2}=0)$ and $\hat\D=\mca G^{-1}-\mca G_0^{-1}$. Ignoring the constant part of the free energy, we perform the expansion up to octic order
\beeq\beal
f&=\fr{2}{|g_{p\txt{VP}}|}(\D_1^2+\D_2^2)+\fr{1}{2}\Tr(\mca G_0\hat\D)^2+\fr{1}{4}\Tr(\mca G_0\hat\D)^4\\&\hspace{11pt}+\fr{1}{6}\Tr(\mca G_0\hat\D)^6+\fr{1}{8}\Tr(\mca G_0\hat\D)^8.
\enal\eneq
The expansion parameter reads
\beeq\beal
\mca G_0\hat\D
=\fr{1}{2}G\sum_ad_a\D_a\tau^3\s^0.
\enal\eneq
Note that the odd-power terms all vanish since $\Tr[(\tau^3)^n]=\Tr\tau^3=0$ for odd $n$'s. The quadratic terms read
\beeq
f^{(2)}
=\lf[\fr{2}{|g_{p\txt{POM}}|}+\fr{1}{2}\Tr(G^2)\ri](\D_1^2+\D_2^2),
\eneq
with the square bracket term $\sim T-T_c$ turning negative below $T_c$, and the quartic terms are
\beeq
f^{(4)}
=\fr{1}{32}\Tr(G^4)(\D_1^2+\D_2^2)^2
\eneq
with positive prefactor. For the sextic terms, we have
\beeq
f^{(6)}
=\fr{1}{96}\Tr\Bigg[G^6\sum_{abcdef}d_ad_bd_cd_dd_ed_f\D_a\D_b\D_c\D_d\D_e\D_f\Bigg].
\eneq
The nonvanishing traces of $d_a$'s are $\Tr(d_1^6)=11/36$, $\Tr(d_1^4d_2^2)=1/36$, $\Tr(d_1^2d_2^4)=1/12$, $\Tr(d_2^6)=1/4$, and the traces of their permutations. This implies
\beeq
f^{(6)}
=\fr{1}{96}\Tr(G^6)(\D_1^2+\D_2^2)^3\lf[\fr{10}{36}+\fr{1}{36}\cos6\t_\D\ri],
\eneq
where $\t_\D=\tan^{-1}(\D_2/\D_1)$ and the prefactor is negative. At octic order
\beeq\beal
f^{(8)}
&=\fr{1}{512}\Tr(G^8)\sum_{abcdefgh}\Tr(d_ad_bd_cd_dd_ed_fd_gd_h)\\&\hspace{11pt}\times\D_a\D_b\D_c\D_d\D_e\D_f\D_g\D_h,
\enal\eneq
the nonvanishing traces of $d_a$'s are $\Tr(d_1^8)=43/216$, $\Tr(d_1^6d_2^2)=1/216$, $\Tr(d_1^4d_2^4)=1/72$, $\Tr(d_1^2d_2^6)=1/24$, $\Tr(d_2^8)=1/8$, and the traces of their permutations. The octic order terms are then obtained as
\beeq
f^{(8)}=\fr{1}{512}\Tr(G^8)(\D_1^2+\D_2^2)^4\lf[\fr{35}{216}+\fr{1}{27}\cos6\t_\D\ri]
\eneq
with positive prefactor. Combining these results, we obtain the Ginzburg-Landau free energy (\ref{eq:freeengpvp}) which is adopted in Sec.~\ref{sec:order}.

\bibliography{Reference}

\end{document}